\documentclass[12 pt, article]{article}
\usepackage[left=1in, right=1in, top=0.75in, bottom=0.75in]{geometry}
\usepackage[utf8]{inputenc}
\usepackage{setspace}
\usepackage{graphicx}
\usepackage{authblk}
\usepackage{sectsty}
\usepackage{gensymb}
\usepackage{hyperref}
\usepackage{parskip}
\usepackage{caption}
\usepackage{titlesec}
\usepackage{wrapfig}
\usepackage{xcolor}
\usepackage[utf8]{inputenc}
\graphicspath{{images/}}
\sectionfont{\fontsize{13}{16}\selectfont}
\subsectionfont{\fontsize{13}{16}\selectfont}

\titlespacing\section{0pt}{12pt plus 2pt minus 2pt}{12pt plus 2pt minus 2pt}
\usepackage[backend=biber, style=nature]{biblatex}
\title{\Large\textbf{Bragg Interferometry of Moiré Superlattices: From Geometric Phase Principles to Atomic Reconstruction}}

\author[1]{Isaac M. Craig}
\author[1,2,3,*]{D. Kwabena Bediako}
\affil[1]{\textit{Department of Chemistry, University of California, Berkeley, CA 94720, USA}}
\affil[2]{\textit{Chemical Sciences Division, Lawrence Berkeley National Laboratory, Berkeley, CA 94720, USA}}
\affil[3]{\textit{Kavli Energy Nanosciences Institute, Berkeley CA, USA, 94720}}
\affil[*]{Correspondence to: bediako@berkeley.edu}

\date{}
\begin{document}
\maketitle
\doublespacing

\section*{Abstract}
The emergence of moiré superlattices formed by twisting and stacking two-dimensional materials has created a need for characterization techniques capable of mapping sub-angstrom atomic reconstruction across micron-scale fields of view. This review surveys a suite of methodologies developed to extract geometric phases in electron microscopy and X-ray spectroscopy, with a primary focus on the interference of overlapping Bragg reflections in the dark field. We provide a comparative analysis of some established techniques, including geometric phase analysis, converged beam electron diffraction holography, and various ptychographic paradigms, culminating in a discussion of Bragg interferometry for measuring interlayer displacement fields and strain in moiré materials. We demonstrate the utility of Bragg interferometry through case studies of twisted bilayer and trilayer graphene as well as transition metal dichalcogenide moiré systems. Special attention is given to the unique advantages of this dark-field interferometric method for probing buried interfaces and encapsulated heterostructures, as well as the inherent challenges of interpreting incomplete phase information in the presence of dynamical scattering. By examining the physical principles underlying these approaches, this review highlights the conceptual similarities and practical trade-offs involved in high-resolution structural mapping of materials in which the interplay between structural relaxation and electronic behavior defines a scientific frontier at the nexus of modern condensed matter physics, nanomaterials engineering, and interfacial chemistry.

\section*{Introduction}
Moiré superlattices of van der Waals (vdW) materials like graphene and transition metal dichalcogenides (TMDs) have provided a transformative platform for condensed matter physics and physical chemistry.\cite{suarez2010flat,bistritzer2011moire,cao2018correlated,cao2018unconventional, chen2019signatures, tran2019evidence,seyler2019signatures,alexeev2019resonantly,jin2019observation,hao2021electric, bai2020excitons,huang2022excitons,mak2022semiconductor, naik2022intralayer,susarla2022hyperspectral,kang2024evidence,yu_tunable_2022,yu2022tuning,zhang_anomalous_2023,vanwinkle2025nanoscale} These “quantum materials” are engineered through the precise stacking of vdW (so-called two-dimensional, 2D) crystals with an azimuthal offset imposed by interlayer twist or small difference in in-plane lattice constant (Figure 1a–d). By introducing a periodic modulation of the atomic environment, these systems create spatially varying potentials that serve as solid-state simulators for correlated electron physics.\cite{tang2020simulation,balents2020superconductivity,yang2022tunable} The twist-dependent electrochemical behavior of these materials\cite{yu_tunable_2022,yu2022tuning,zhang_anomalous_2023,vanwinkle2025nanoscale} also sparked discovery of the electronic contribution to the reorganization energy in the Marcus electron transfer framework.\cite{CoelloEscalante2024TBG,Maroo2026electronic} However, the idealized “rigid” moiré lattice is a fiction; in practice, the competing forces of interlayer stacking energies and intralayer elasticity drive a process of local atomic reconstruction (Figure 1e,f).\cite{Parkinson1991periodic,alden2013strain,woods2014commensurate,wijk2015relaxation,dai2016twisted,nam2017lattice,tadmor2018structural,carr2018relaxation,yoo2019atomic,weston2020atomic,kazmierczak2021strain,nakatsuji2023multi} Even minor variations in this structural relaxation and attendant strain fields fundamentally alter electronic behavior\cite{nam2017lattice,carr2018relaxation,yoo2019atomic} and dictate interfacial electron transfer properties.\cite{yu_tunable_2022,zhang_anomalous_2023,vanwinkle2025nanoscale} Moiré superlattices are particularly susceptible to moiré scale disorder\cite{uri2020mapping,wilson2020disorder,lau2022reproducibility} and heterostrain,\cite{huder2018electronic,bi2019designing,kerelsky2019maximized} with minuscule atom scale distortions becoming magnified on the moiré lengthscale. Consequently, developing characterization techniques that are capable of mapping what are effectively sub-angstrom distortions over micron sized samples containing ca. 10,000s of moiré periods is a central priority for the field of moiré quantum matter.

\begin{figure}
\begin{center}
\includegraphics[width=10cm]{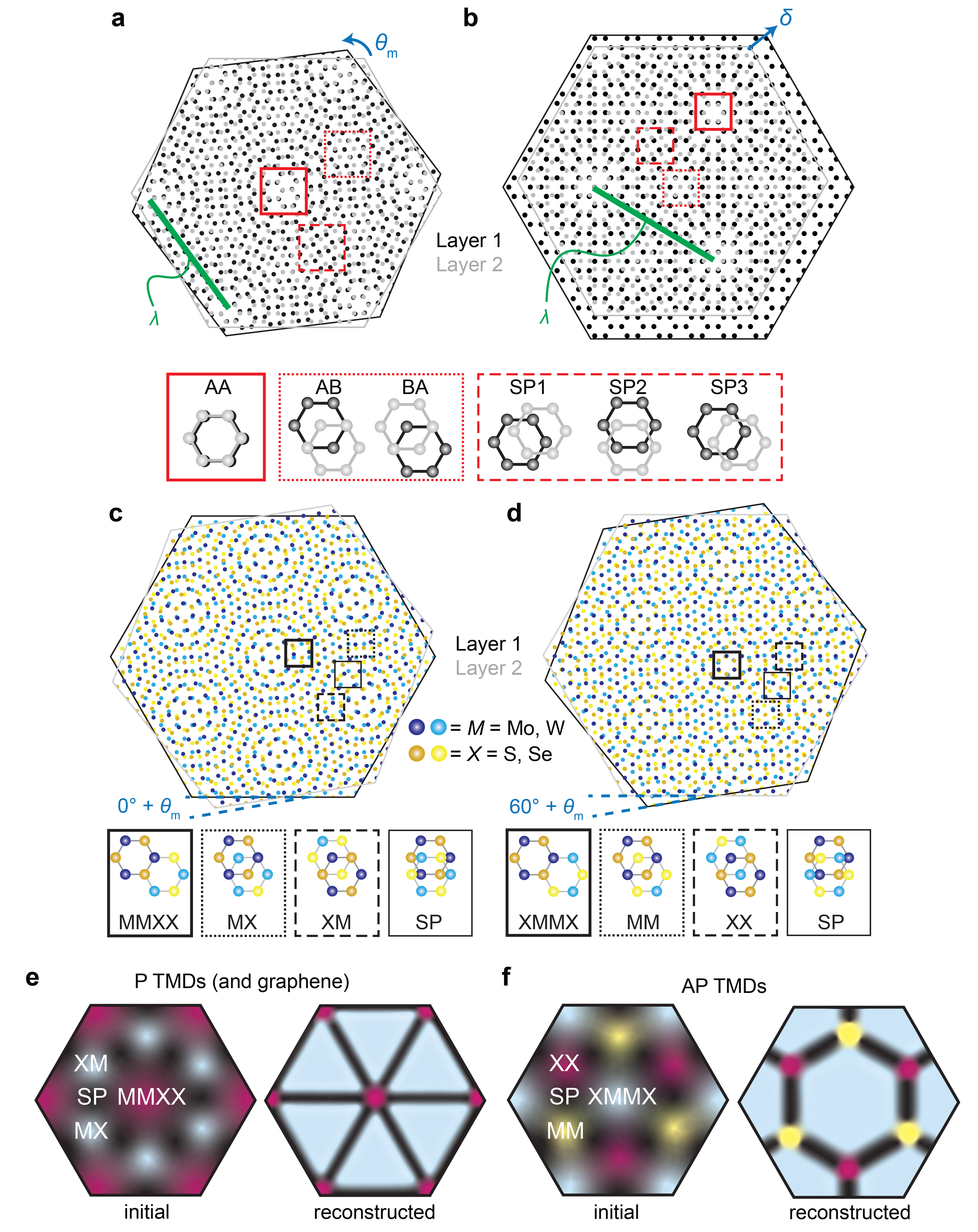}
\end{center}
\caption{ \textbf{Bilayer vdW moiré superlattices.} (a, b) Rigid lattice models of moiré superlattices from (a) interlayer twist, $\theta_m$, and (b) lattice constant difference, $\delta$, in prototypical triangular lattice bilayers. The moiré superlattice constant, $\lambda$, is depicted. (c, d) Rigid lattice models of (c) parallel, P (near 0$^{\circ}$), and (d) anti-parallel, AP (near 60$^{\circ}$), twisted bilayer TMD configurations. Below each set of rigid models in a–d are the corresponding top-down views of the distinct sets of interlayer stacking orders found in each moiré. (e, f) Schematics depicting the formation of expanded low energy stacking regions due to atomic reconstruction of moiré superlattices. Similar relaxed structures showing triangular domains are expected in P TMDs and graphene (e) whereas a honeycomb-like structure is expected in AP TMDs (f). These domain configurations are based on the different stacking orders present in each superlattice, their spatial arrangement, and their relative energies.} 
\end{figure}

Characterizing these buried interfaces requires a choice of probe that balances spatial resolution with signal sensitivity. While X-ray ptychography\cite{hruszkewycz2017high,pfeiffer2018x,kim2018three,takahashi2013bragg} has long been a gold standard for high-strain sensitivity in bulk materials due to the deep penetration of photons, electron-based probes offer the significantly higher scattering cross-sections and sub-nanometer real-space resolution necessary to resolve the nanosized domains and sharp domain walls characteristic of marginal twist angles.\cite{dai2016twisted,tadmor2018structural,yoo2019atomic,kazmierczak2021strain} Within the suite of four-dimensional scanning transmission electron microscopy (4D-STEM)\cite{ophus2014recording,ophus2019four} techniques, several paradigms have emerged to quantify such nanosized features. These range from comprehensive phase-retrieval methods like electron ptychography,\cite{humphry2012ptychographic,pennycook2015efficient,gao2017electron,jiang2018electron,chen2020mixed,yang2024local,pelz2017low,chen2021electron} which seeks to reconstruct the full electron wavefunction from diffraction intensities acquired at overlapping probe positions, to real-space approaches such as geometric phase analysis (GPA)\cite{hytch1998quantitative,Pasztor2019Holographic,Pasztor2021multiband,pasztor2024delusive,ke2022moire} that map lattice displacements and strain.

In this review, we introduce the conceptual bases for some of these characterization methodologies, with a particular focus on Bragg Interferometry (BI),\cite{kazmierczak2021strain,van2023rotational,craig2024local,craig2024considerations,vanwinkle2025nanoscale} placing it in the context of the broad family of techniques which proceeded it. Indeed, the foundational concept—exploiting the interference fringes from overlapping Bragg reflections to determine the stacking orientation of layered materials—is certainly not new, appearing often across a range of techniques in both electron microscopy and X-ray spectroscopy. In brief, we define BI as a constrained analogue to dark-field ptychography that extracts structural details of moiré materials (specifically, interlayer displacement and intralayer strain resulting from atomic relaxation) by fitting the intensity interference of overlapping Bragg disks to analytical functional forms. Unlike electron ptychography, which seeks to reconstruct the entire probe wavefunction, BI offers a ``coarse-grained” picture that is computationally efficient and distinctively suited for mapping large-area moiré reconstruction.

Following a concise overview of existing ptychographic and holographic techniques, we discuss the utility of Bragg interferometry through a series of case studies. We explore the insights gained from BI into structural landscapes of twisted bilayer graphene (tBG) \cite{kazmierczak2021strain}, twisted TMD bilayers \cite{van2023rotational}, and twisted trilayer graphene (tTG) \cite{craig2024local}, illustrating how BI disentangles the complex interplay between atomic reconstruction and the exotic physical and chemical properties of these quantum materials. First, owing to the conceptual connection between the electron-based techniques discussed here and analogues in X-ray spectroscopy, it is useful to begin with a brief discussion of some relevant differences between electron and X-ray probes. 

\section*{Comparing X-ray and electron probes}

The first prototypical electron microscope was developed in 1931, \cite{ruska1931magnetische} nearly two decades after the first X-ray spectrometers. Accordingly, the field of electron microscopy has frequently adapted methodologies originally pioneered for its older sibling, translating the language of photon diffraction into the regime of charged-particle optics. However, several fundamental distinctions between the two probes necessitate distinct experimental considerations and analytical frameworks. Arguably, the primary motivations for the development of electron-based probes were the significantly shorter wavelengths and stronger interaction cross-sections inherent to electrons. While hard X-rays typically possess characteristic wavelengths on the order of 0.1 nm, the relativistic, voltage-dependent wavelength of a 300 kV electron is roughly 0.02 Å. This allows for diffraction-limited resolution comparable to gamma radiation, yet without the catastrophic ionization damage associated with high-energy photons, as electrons primarily undergo Rutherford scattering. Crucially, the elastic scattering cross-section for electrons is approximately five to six orders of magnitude larger than that for X-rays, making electrons uniquely sensitive to the subtle potentials of atomically thin 2D materials. 

This high sensitivity comes with a trade-off in the complexity of the scattering event. Because electrons interact so strongly with the electrostatic potential of the sample, multiple scattering (dynamical diffraction) becomes a significant factor even in relatively thin specimens, whereas X-ray scattering is typically well-described by the simpler kinematical approximation. Furthermore, the electromagnetic coils used to focus electrons are considerably more prone to geometric and chromatic aberrations than the optical elements used for X-rays.\cite{scherzer1936,scherzer1949theoretical} While the advent of aberration correction has pushed electron microscopy into the sub-angstrom regime, X-ray optics still maintain a superior ability to preserve the coherence and wavefront of the probe over long distances.

Practical differences in detection and penetration depth further distinguish the two fields. X-rays possess a much greater penetration depth, allowing for the non-destructive characterization of buried interfaces in bulk crystals or within complex in situ environments. Electrons, by contrast, are essentially surface or thin-film probes, making them the natural choice for the study of isolated vdW heterostructures. Additionally, while electron detectors have seen a revolution in direct-electron counting (offering nearly noiseless detection and high dynamic range)\cite{faruqi2011electronic,levin2021direct} X-ray detection still faces challenges in matching these signal-to-noise ratios at the same spatial frequencies. Ultimately, a direct comparison of these techniques reveals that while X-ray ptychography remains the standard for bulk strain mapping, the unique combination of high scattering cross-sections and sub-nanometer real-space resolution renders electron-based interferometry the superior tool for resolving the delicate atomic reconstructions found in marginally twisted moiré systems. In light of these connections, the discussion of common techniques benefits from some comparisons between the approaches of both fields, as presented in what follows.

\section*{The ptychographic framework and its variants}

Many detailed reviews of ptychography\cite{Rodenburg2019,pfeiffer2018x,zheng2021concept,rodenburg2025ptychography,clark2025electronptychography,miao2025computational,wang2023optical} and its history are available. The technique is outlined here as context for understanding the broad family of methods that may extract sample structure from overlapping Bragg reflections and their history. The data used in electron ptychography are often a 4D-STEM \cite{ophus2014recording,ophus2019four} dataset. Figure 2 compares the measurement setup for conventional TEM (Figure 2a) with that of converged beam electron diffraction, CBED (Figure 2b). A 4D-STEM measurement is a two-dimensional collection (over scan location) of the two-dimensional CBED patterns (Figure 2c).  

\begin{figure}[tbhp]
\begin{center}
\includegraphics[width=11cm]{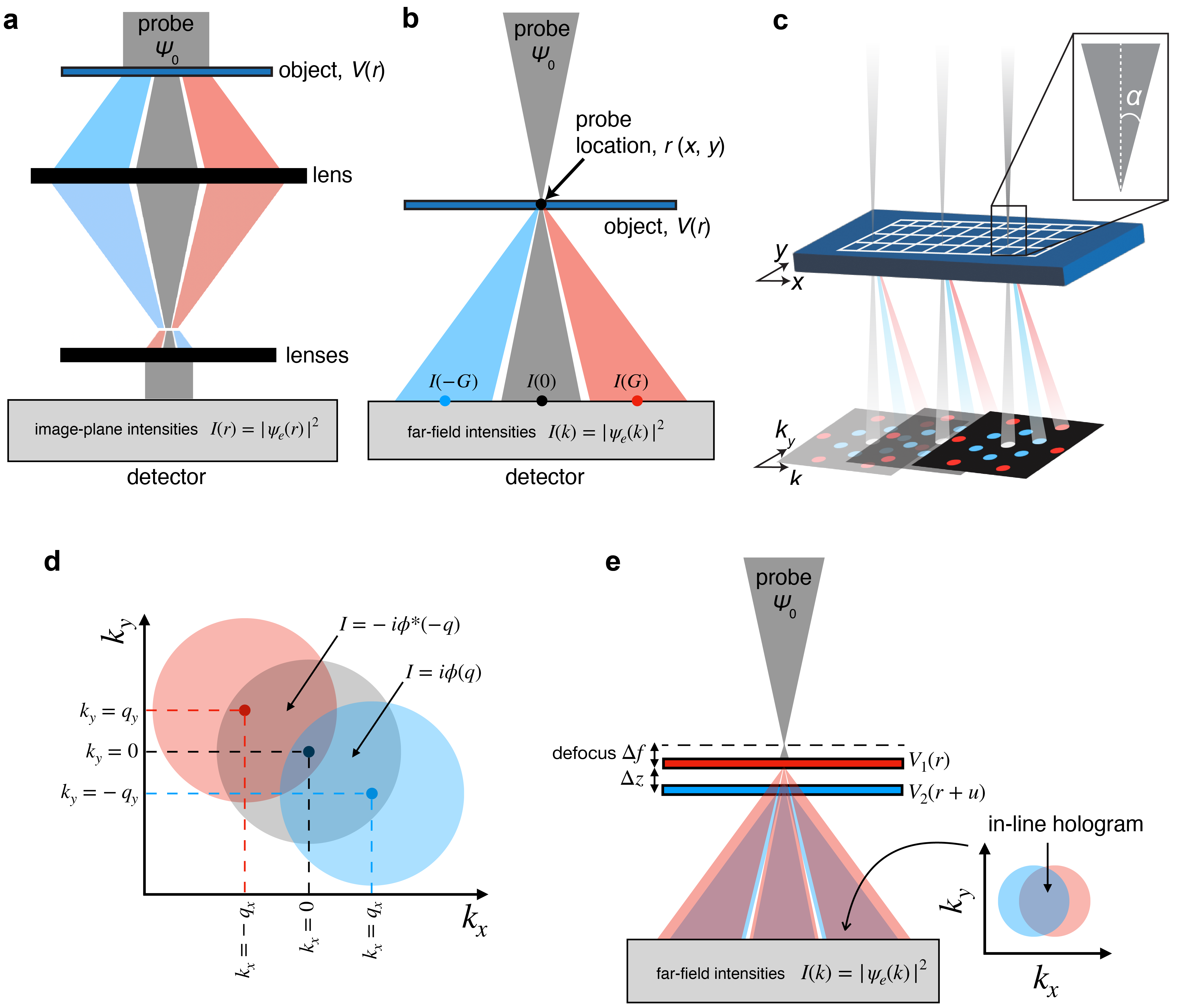}
\end{center}
\caption{ \textbf{Imaging Modes.} (a,b) Schematics illustrating the detector set-up for conventional TEM (a) and CBED (b). (c) Schematic for 4D-STEM, where a collection of CBED patterns are collected across a range of probe positions $r$. The convergence angle, $\alpha$, of the beam controls the diffraction disk diameter. (d) Schematic illustrating the intensities $I(k,q)$ (Equation 3) for an aberation-free probe. (e) Schematic illustrating the setup for in-line CBED holography and BI.} 
\end{figure}

Ptychography in essence involves correcting for poorly focused or even absent lenses with computational optimization. This is achieved by over-collecting data for use in algorithms which aim to determine both the sample scattering potential (also referred to as the object $V(r)$) and the functional form of the potentially highly aberrated probe, and necessarily involves various either redundancies in information or direct measurement of interference fringes to circumvent the phase problem (namely, the loss of phase information when measuring $|\psi|^2$).\cite{Rodenburg2019} Ptychography is now used to refer to any technique that exploits some flavor of optimization to extract both the (aberrated) wavefunction of the incident probe wave $\psi_0$ and the sample potential or object $V$ via a series of measurements of exit wave intensity $I = |\psi_e|^2$. The relationship between the probe $\psi_0$ and exit wave $\psi_e$ is represented often by the transmission function $T$ or phase shift $\phi \propto V$, encoding the details of the scattering mechanism, and grows in complexity in the face of inelastic events and multiple scattering (terms with a higher order of the interaction parameter in the Born series), 
\begin{equation}
    \psi_e(r) = T[\psi_0(r)] =  e^{i\phi(r)} \psi_0(r) 
\end{equation}

The original use of the term ptychography was by Hoppe and Hegerl to describe a method for extracting the phase of Bragg reflections from the interference fringes present in overlapping Bragg peaks, with conjugate ambiguity resolved by monitoring changes with lateral sample movement.\cite{Hoppe1969,hegerl1970dynamische} Despite the approach for which it was originally coined, modern development and use of ptychography often bears little resemblance to this original concept. Most ptychographic techniques circumvent the phase problem through collecting data covering a range of lateral sample shifts, in which the equivalence of oversampled regions is used as a constraint.\cite{Rodenburg2019} The technique benefits from the increased ability to adjust the lateral shifts of a sample, in comparison to, for example, a focal series. Some techniques however exploit the interference fringes for which ptychography owes its name. One example is single sideband (SSB) ptychography.\cite{rodenburg1993experimental}

\subsection*{Single sideband ptychography} 
SSB ptychography \cite{rodenburg1993experimental} is an analytical ptychography method, wherein the transmission function is obtained from analytical expressions directly rather than via a more numerically intensive optimization procedure to obtain both the object and probe simultaneously. Analytical ptychography methods often obtain the object under the assumption of an ideal probe, but include strategies to account for the approximate impact of aberrations. In the case of SSB ptychography, the phase shift $\phi(r)$ (and therefore object $V \propto \phi$) is obtained from the intensities of overlapping Bragg reflections. To clarify notation, the phase shift $\phi(r)$ induced in the probe $\psi_0$ by interaction with the object is defined as follows. The first-order Taylor expansion of phase $\phi(r)$, valid for weakly scattering samples, is referred to as the weak phase object approximation,
\begin{equation}
    \psi_e(r) = e^{i\phi(r)} \psi_0(r) \approx \psi_0(r) + i\phi(r)\psi_0(r).
\end{equation}
We begin by noting that the 4D-STEM data, or the collection of far-field CBED intensities $I(k,r) = |\psi_e(k,r)|^2$ acquired over a collection of probe locations $r$, has a Fourier transform that takes the following form: \cite{rodenburg1993experimental}
\begin{equation}
    I(k, q) = FT[I(k,r)] = |\psi_0(k)|^2\delta(q) + i \psi_0^*(k) \psi_0(k-q) \phi(q) - i \psi_0(k) \psi_0^*(k+q) \phi^*(-q).
\end{equation}

In the above, $\psi_0(k)$ is the reciprocal space probe wavefunction, represented in terms of the aperture $A(k)$ and aberration function $\chi(k)$ as $\psi_0(k) = A(k)e^{i\chi(k)}$. $\delta$ is the Dirac-delta function, and $\phi$ is the Fourier transform of the phase shift map. This corresponds to three Bragg disks (with a shape, often circular, set by $A(k)$) centered at $0$, $q$, and $-q$ (see Figure 2d). The double-overlap regions are referred to as side-bands and have intensities of $i\phi(q)$ and $-i\phi^*(-q)$ when the probe is aberration-free (when $\chi(k)=0$). Triple overlap regions are here neglected, and contain no phase information from the cancellation of the last two terms in equation 2. The phase shift $\phi(r)$ is then obtained from the inverse Fourier transform of $\phi(q)$. Aberrations can also be accounted for approximately in this framework. It should also be noted that the obtained $\phi(q) \in (\pi, -\pi]$ contains discontinuous jumps and requires the use of a phase unwrapping algorithm. While this technique appears to extract phase information from only overlapping Bragg reflections in the dark-field, it is distinct from some subsequently discussed methods in that it considers all $q$ choices in the Fourier transformed $I(k, q)$, and therefore all $k$. The bright-field information (and therefore complete geometric phase information) is therefore included. 

\subsection*{Dark-field X-ray ptychography} 

While the inclusion of dark-field information is advantageous \cite{humphry2012ptychographic} in electron ptychography, completely obscuring the bright-field region prevents full reconstruction due to the loss of phase information discussed throughout this review. The ptychographic technique that comes closest to using only dark-field information is dark-field X-ray ptychography.\cite{suzuki2016dark} Dark-field X-ray ptychography was motivated by challenges in obtaining a detector dynamic range sufficient to simultaneously acquire intensities at low and high scattering vectors for X-rays: a problem less pressing in electron microscopy from the extremely high dynamic range of electron microscope pixel array detectors. \cite{tate2016high} This problem is addressed in dark-field X-ray ptychography through placing a cylindrical reference object between the probe source and object, and acquiring only dark-field intensities. The reference object is introduced to create in-line holograms within the dark-field to acquire the additional phase information lost from obscuring the central bright-field image. \cite{suzuki2016dark} In a bilayer material, one layer acts as a reference object so the overlapping Bragg reflections contain in-line holograms in a similar manner. \cite{latychevskaia2018convergent} The discussed differences between X-ray and electron probes, however, imply that there will be far fewer strongly excited features in a CBED pattern so there is an expected relative loss of phase information. Bragg interferometry, in essence, involves analytically interpreting these in-line holograms from few-layered materials, assuming negligible multiple scattering. 

\section*{Related alternative mapping solutions}
\subsection*{Geometric phase analysis}

It should be noted that this concept of extracting geometric information from interference fringes appears ubiquitously outside of pytchography as well. The prior section discussed how the far-field intensity distribution $I(k)$ acquired in a diffraction or converged beam mode (see Figure 2b) can be analyzed to assess $V$, however a similar logic can be applied to interpret the bright-field image $I(r)$ (or rather, its Fourier series). \cite{hytch1998quantitative} This bright-field intensity $I(r)$ from scattering off a perfect crystal can be expanded as follows, where G are the reciprocal lattice vectors of the crystal and $A_G e^{iP_G}$ are the (complex) Fourier coefficients. The fact that $I(r)$ is real was used. 
\begin{equation}
    I(r) = \sum_{G} A_G e^{2\pi i \textbf{G} \cdot \textbf{r} + iP_G} = A_0 + \sum_{G>0} 2A_G \cos{(2\pi \textbf{G} \cdot \textbf{r} + P_G)}
\end{equation}
The broadening of the intensity away from a perfect collection of exponentially localized points within an imperfect crystal can be introduced by allowing $A_G$ and $P_G$ to depend on $r$.
\begin{equation}
    I(r) = \sum_{G} A_G(r) e^{2\pi i \textbf{G} \cdot \textbf{r} + iP_G(r)} = A_0(r) + \sum_{G>0} 2A_G(r) \cos{(2\pi \textbf{G} \cdot \textbf{r} + P_G(r))}
\end{equation}
Placing masks around the peak at $\textbf{G}$ in the Fourier transform of $I(r)$ and performing an inverse Fourier transform will extract the complex image $A_G(r) e^{2\pi i \textbf{G} \cdot \textbf{r} + i P_G(r)}$, from which the local phase image $P_G(r)$ can be obtained. This is different from Bragg filtering to obtain the real image $2A_G(r) \cos{(2\pi \textbf{G} \cdot \textbf{r} + P_G(r))}$ from both $\pm \textbf{G}$, which introduces a phase ambiguity complicating $P_G(r)$ extraction. This distinction will be discussed again later as it is one complication of techniques that aim to collect phase information from (real) measurements of diffraction space intensities instead. Strain in the underlying crystal can be introduced in the form of an (infinitesimal) displacement field $\textbf{u}(r)$ that maps the idealized atomic positions $\textbf{r}$ onto their actual locations $\textbf{r'}(r) = \textbf{r} + \textbf{u}(r)$. This will yield a change in phase of $P_G(r) = 2\pi \textbf{G} \cdot \textbf{u}(r)$ for the local value of $\textbf{u}(r)$. The $P_G(r)$ of multiple non-parallel $\textbf{G}$ choices must be used to extract the total $\textbf{u}(r)$ from its $\textbf{G} \cdot \textbf{u}(r)$ projections, and the signal-to-noise ratio can be increased with a larger set of phase images $P_G(r)$. The strain $\nabla \textbf{u}(r)$ can be obtained similarly with use of the following expression, which avoids the need to determine a smoothly varying value of $\textbf{u}(r)$ from the phase-wrapped $P_G$ within the range of 0 to $2\pi$. \cite{hytch1998quantitative}
\begin{equation}
    \nabla P_G(r) = \Im(e^{-iP_g(r)} \nabla e^{iP_g(r)}) = 2\pi\textbf{G} \cdot \nabla\textbf{u}(r)
\end{equation}

Assuming that variations in the phase image $P_G$ are entirely driven by this term proportional to $\textbf{G} \cdot \textbf{u}$ involves neglect of $O((\textbf{G} \cdot \textbf{u})^2)$ and higher terms. This is associated with assuming a fixed reciprocal lattice reference $\textbf{G}$ and is related to the broader family of lock-in techniques of which GPA is a member. \cite{de2022imaging}

In summary, the strain of a sample is unambiguously determined from the Fourier spectrum of a high-resolution bright-field image and loss of phase information is not a complication. One way to understand this is by noting that the portion of the exit wave $\psi_e(\textbf{G})$ that is associated with scattering to Bragg reflection at $\textbf{G}$ carries useful phase information, but such phase information is lost in the intensity measurement of an isolated Bragg peak $|\psi_e(\textbf{G})|^2$. It is only when multiple Bragg disks overlap that interference fringes encoding phase information will be present. As the bright-field image contains superpositions of the $\psi_e$ components associated with all $\textbf{G}$, all of these fringes are present within the image, and the complete complex image and associated phase can be obtained. This is not the case in the dark-field, where at most a few $\textbf{G}$ components will overlap. Such fringes from overlapping Bragg peaks can however can still be interpreted, as is done in CBED holography \cite{latychevskaia2018convergent} and interferometry. \cite{kazmierczak2021strain}  

\subsection*{CBED holography}

Holography, introduced by Gabor in 1948, \cite{gabor} involves the extraction of sample structure from the interference pattern (hologram) created from the superposition of multiple split waves, such as a probe and reference laser beam in the optical regime. In the context of electron microscopy, this could involve the interference pattern created by an overlapping reference probe and the exit wave. Common variants include off-axis holography, where the beam is split into two parallel probes, one of which interacts with the object, before recombination. Another common alternative geometry is in-line holography, where a single probing beam is used, and the hologram is formed from interference between the scattered and transmitted portions of the probe. In a bilayer material, holograms can also be formed from interference between the portion of the probe that scattered strongly off of each individual layer, as is the case in in-line CBED holography. \cite{latychevskaia2018convergent}

Within CBED and 4D-STEM, the diffraction spots become large Bragg disks (Figure 2b), and the intensity variation within the disks is accessible. When multiple Bragg disks overlap, interference fringes appear that can be analyzed to determine geometric information. It should be noted here that now a single CBED pattern is analyzed, in contrast to the Fourier spectrum of a bright-field image or the collection of 4D-STEM data as in GPA and SSB ptychography. The intensity variation $I(k,r_0)$ within the in-line hologram (Figure 2e) can be explicitly written out for interaction with a bilayer material, where $\textbf{u}(r_0)$ and $\Delta z(r_0)$ are the local out-of-plane and in-plane inter-layer distances at fixed probe position $r_0$. It is as follows:
\begin{equation}
    I(k, r_0) \propto \cos(c_1\Delta z(r_0) - 2 \pi \textbf{k} \cdot \textbf{u}(r_0) + \Delta \gamma).
\end{equation}
 $c_1$ is a constant equal to $\pi \lambda/a^2$ where $\lambda$ is the electron wavelength for the probe and $a$ the lattice spacing for a given reflection. $\Delta \gamma$ is an offset that corresponds to redefining the origin of the displacement field. For instance when $\Delta \gamma=0$, $\textbf{u}$ measures the extent of deformation away from perfect out-of-plane alignment or AA-stacking. Small deformations from another stacking order may instead be chosen. 

From equation 7, it is apparent that the spacing of fringes within the hologram is related to the separation between monolayers $\Delta z$. Additionally, relative in-plane displacements of one layer relative to the other (i.e., the local atomic stacking) shifts the fringes. In the referenced work, \cite{latychevskaia2018convergent} this framework was used to determine the inter-layer spacing $\Delta z$ and stacking order within the illumination region of a single CBED pattern, and benefits from low dose. Additionally, this approach uses exclusively dark-field information and can therefore be applied to encapsulated materials as even strongly scattering bystander layers can be ignored provided that they are sufficiently misoriented (i.e., that Bragg disks associated with scattering off of them are at sufficiently different reciprocal lattice vectors to avoid overlap). 

\section*{Bragg interferometry}

While CBED holography benefits from the ability to analyze a single CBED pattern, one complicating factor is the inability to see multiple fringes in the limit of a well-focused probe and small sample interaction area for high resolution. Instead, the expressions can be interpreted in the limit where the fringe spacing that encodes $\Delta z$ is obscured and only the fringe origin offset values are seen. The collection of a full 4D-STEM dataset of variable $r$ then allows for $\textbf{u}(r)$ to be extracted. \cite{kazmierczak2021strain, zachman2021interferometric} The overlap intensity can be derived starting from the weak phase object approximation for scattering off of the object $V_1(\textbf{r}-\textbf{u}/2) + V_2(\textbf{r}+\textbf{u}/2))$. 
\begin{equation}
I(\textbf{k}) = |\psi_0(\textbf{k})|^{2} + 
\sigma^2 \sum_\textbf{g} |\psi_0(\textbf{k}-\textbf{g})|^2 ( |V_1(\textbf{g})|^2  + |V_2(\textbf{g})|^2 +
2 Re( V_1(\textbf{g})V_2(\textbf{g})^* e^{2i\pi \textbf{g} \cdot u}) )
\end{equation}

The resultant CBED intensities $I(\textbf{k})$ above were obtained assuming that Bragg disks within a given layer do not overlap. This leads to the following relationship in equation 8 for overlap intensity $I_j$ at reflection $\textbf{g}_j$. $C(g) \propto \Im(V_1(g)V_2(g)^*)$ is only present for heterobilayers ($V_1 \neq V_2$) that lack in-plane centrosymmetry, such that the Fourier-space projected electrostatic potentials $V_i$ are complex. \cite{van2023rotational} Additionally, $A(g)$ is nonzero only for $V_1 \neq V_2$. Extensions to multiple overlap regions are straightforward and contain a sum of all the following terms for each total displacement $\textbf{u}_{ij}$ between layers $i$ and $j$. 
\cite{craig2024local} The displacement field origin can be chosen to eliminate $C(g)$ in some materials and is equivalent to adjusting $\Delta \gamma$ in equation 7.

\begin{equation}
I_j(r) = A_j + B_j cos^2(\pi \textbf{g}_j \cdot \textbf{u}(r)) + C_j cos(\pi \textbf{g}_j \cdot \textbf{u}(r)) sin(\pi \textbf{g}_j \cdot \textbf{u}(r)) 
\end{equation}

The displacements $\textbf{u}(r)$ can then be obtained from fitting $I_j(r)$ to this expression (Figure 3). The impact of the out-of-plane distance between layers $\Delta z$ is neglected as the impact of free-space propagation between the two layers is not included in equation 8, unlike in the derivation of equation 7. We found the slight shift from $\Delta z$ is too small to obtain a reliable estimate of the interlayer distance in practice when using the imaging conditions in the referenced works. \cite{kazmierczak2021strain, van2023rotational, craig2024local} The small impact of $\Delta z$ on equation 1.10 can be corrected for by averaging the intensities within the Friedel pair Bragg disks at $\pm \textbf{g}$ for which the shifts are equal and opposite (to first order). 

\begin{figure} [btp]
\begin{center}
\includegraphics[width=9cm]{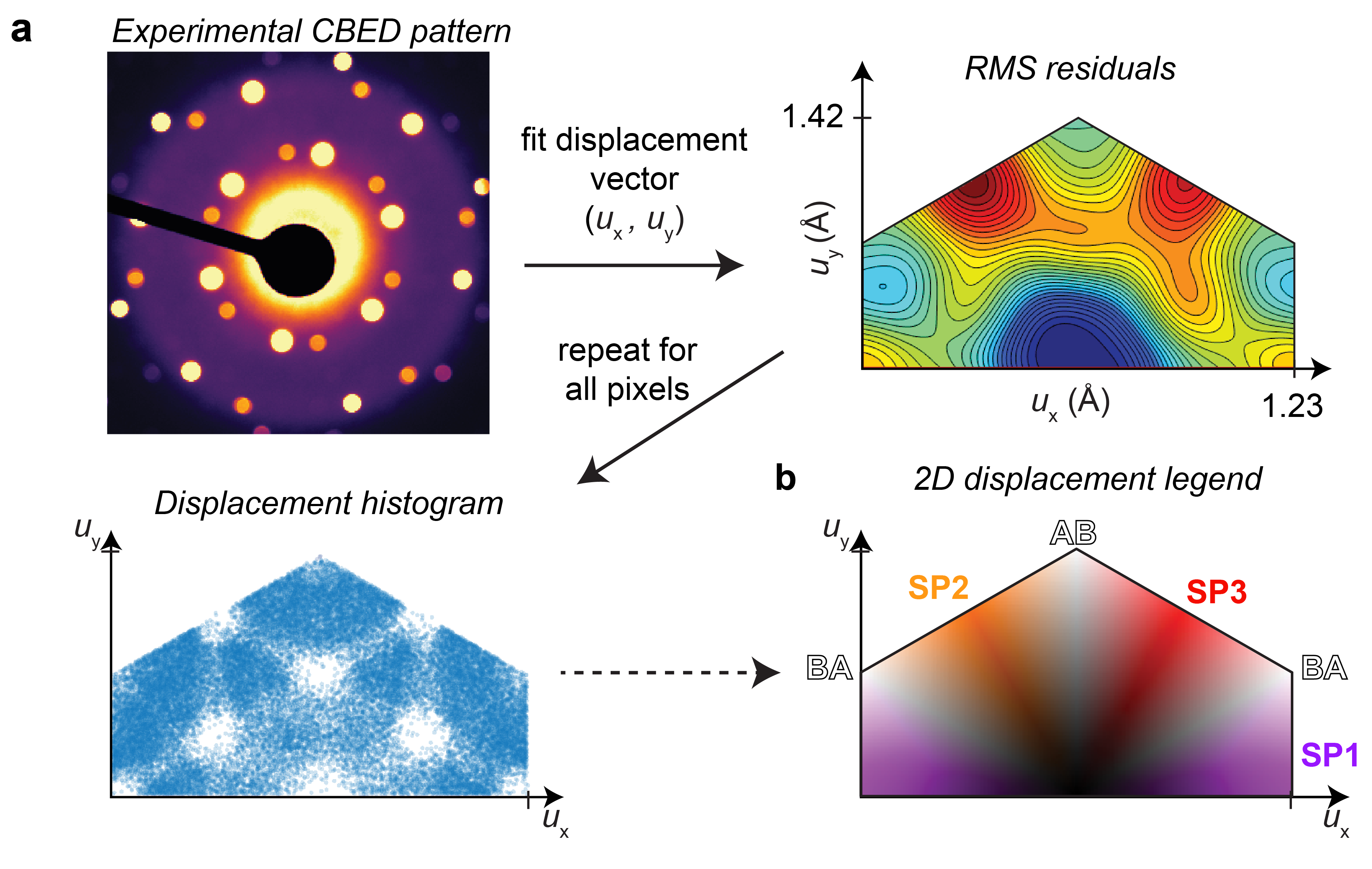}
\end{center}
\caption{  \textbf{Bragg interferometry fitting routine.} (a) Method for extracting local displacement vectors $\mathbf{u}(r)=(u_x, u_y)$ from Bragg disk intensities, $I_j$. (b) 2D colorization scheme used to produce displacement maps from the fitted displacement vectors. Reproduced with permission from reference \cite{kazmierczak2021strain}.} 
\end{figure}

One complication of this approach, and any such approach that uses exclusively dark-field information, is the loss of phase information contained in the bright-field. BI can be thought of in the context of GPA as providing the change in phase $P_G = \pi \textbf{G} \cdot (\textbf{u}_{top}(r) - \textbf{u}_{bottom}(r))$ (where a single $\textbf{G}$ being the average of both layers is assumed). As BI only measures the difference in geometric phase between two or more layers, it relies on an assumption for the average lattice constant and that the displacement in equally partitioned between both layers. The average lattice constant can be obtained approximately from the Bragg peak locations, but a precise measure of the exact Bragg peak locations for each layer is often not possible for the moiré materials investigated with Bragg Interferometry (BI), or else these alone would provide sample structure and avoid the need for BI entirely. Another key distinction is that only the absolute value of the phase image is obtained so equation 5 cannot be used and unwrapping is needed. BI, like the other dark-field methods mentioned, benefits from the ability to probe encapsulated materials. While Fourier filtering could, in principle, provide the same advantage in GPA, such manual removal of bystander layers would have significantly worse signal-to-noise ratio than a dark-field method, prohibitively so when one of these bystander layers scatters strongly.

\subsection*{Case studies and applications}
The prior sections discussed the extraction of displacement fields from overlapping Bragg reflections and the conceptual relationship between BI and related approaches. It is useful to briefly review several representative applications of BI to moiré materials, as these examples illustrate both the structural information accessible through the technique and the broader role of atomic reconstruction in determining electronic properties.

\begin{figure}
\begin{center}
\includegraphics[width=11cm]{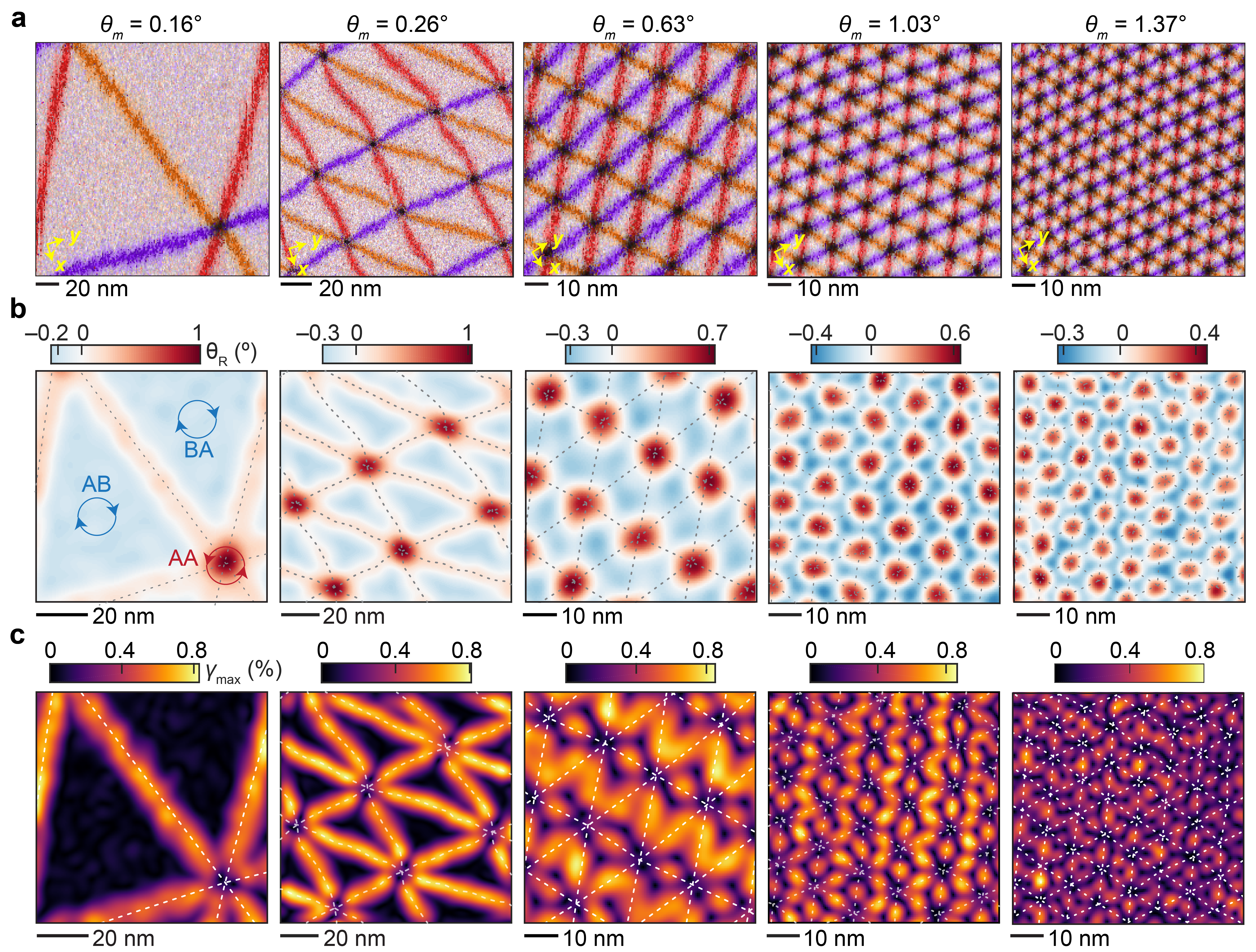}
\end{center}
\caption{  \textbf{Mapping strain fields in tBG.} (a) Displacement field maps for tBG at various moiré twist angles, $\theta_m$, using the colorization scheme in Figure 3b. (b) Maps of combined reconstruction rotation, $\theta_R$, of both layers at each pixel for the regions shown in c. (c) Corresponding maps of average principal shear strain, $\gamma_{max}$, per graphene layer at each pixel. Reproduced with permission from reference \cite{kazmierczak2021strain}.} 
\end{figure}

To date, BI has primarily been applied to twisted bilayer graphene (tBG), twisted transition metal dichalcogenide (tTMD) heterobilayers, and twisted trilayer graphene (tTG), systems in which structural relaxation is expected to strongly modify the local stacking configuration and therefore the low-energy electronic structure. These studies highlight how direct measurements of displacement fields can provide information that is difficult to obtain from diffraction peak positions or real-space imaging alone.

\subsubsection*{Twisted bilayer graphene}

The first realization of BI was applied to tBG with twist angles below approximately $2^\circ$, where lattice relaxation plays an increasingly important role in determining the geometry of the moiré superlattice. \cite{kazmierczak2021strain} In this regime, the energetics of the system favor the expansion of low-energy AB and BA stacking regions and the contraction of higher-energy AA regions, resulting in substantial deviations from the rigidly twisted geometry often assumed in continuum descriptions.\cite{bistritzer2011moire} Scanning tunneling microscopy \cite{kerelsky2019maximized} and dark-field TEM \cite{yoo2019atomic} provided experimental evidence for reconstruction, validating previous simulations that had established the theoretical basis for atomic reconstruction in tBG\cite{dai2016twisted,jain2017structure,tadmor2018structural,carr2018relaxation,}. However, direct experimental measurements of these reconstructed materials were limited, and strain measurements had been limited to determinations of 1D strain\cite{alden2013strain} and heterostrain\cite{kerelsky2019maximized}. Key challenges for imaging tBG moirés and determining the relevant strain tensors associated with relaxation were (1) the inherent constraint of imaging samples comprised of light atoms and (2) the near-unavoidable difficulty of imaging interfaces buried in encapsulating hexagonal boron nitride (hBN), which is typically used in sample fabrication.

By fitting the overlap intensities associated with multiple graphene Bragg reflections, BI was used to obtain the local displacement field $\mathbf{u}(r)$ throughout the moiré lattice (Figure 4a). This enabled direct visualization of the structural relaxation that accompanies decreasing twist angle. Rather than behaving as a uniformly twisted bilayer, tBG was also shown to contain significant spatial variations in both local twist angle, $\theta_m$ and heterostrain $\varepsilon_H$. These fluctuations occur on length scales much shorter than typical overall device dimensions and indicate that disorder in moiré materials—previously only considered in terms of ``long-range" $\theta_m$ disorder, fabrication imperfections, or contamination\cite{uri2020mapping}—often originates from highly local structural variations of both $\theta_m$ and $\varepsilon_H$ from one moiré domain to another.\cite{kazmierczak2021strain}

Taking the gradient of local displacement fields (Figure 4a) enabled the determination of the 2D strain tensor at each probe position, providing a high-fidelity visualization of both reconstruction rotation (Figure 4b) and shear strain (Figure 4c).\cite{kazmierczak2021strain} This detailed strain analysis revealed that atomic reconstruction in tBG proceeds through two distinct regimes, where the persistence of flat-band physics is intimately linked to the specific mechanics of structural relaxation. Near the first magic angle of $1.1^\circ$, the reconstruction is dominated by local AA-site rotations. This structural motif serves to isolate the flat bands from dispersive high-energy states, effectively stabilizing the conditions necessary for correlated electron phases.\cite{torma2021superfluidity} This discovery was pivotal, as it challenged the earlier presumption that lattice relaxation was a universally detrimental perturbation; instead, around $1.1^\circ$, relaxation is a constructive force that promotes flat-band formation. However, as the twist angle decreases toward $0.5^\circ$, a structural transition occurs. In this regime, AB/BA counter-rotations become the dominant reconstruction motif, frustrating the delicate symmetry required for the emergence of the other predicted magic angles around $0.5^\circ$, $0.35^\circ$, $0.24^\circ$, and $0.2^\circ$.\cite{bistritzer2011moire} This shift in the relaxation mechanism, from AA-dominant to AB-counter-rotational, provides a structural basis for why the flat bands and their associated correlated physics degrade at ultra-low twist angles, despite theoretical predictions based on rigid lattice models.

The geometric consequences of lattice relaxation in tBG extend beyond electronic band structure, fundamentally dictating the kinetics of interfacial electron transfer (ET) in electrochemical systems \cite{yu_tunable_2022,yu2022tuning,zhang_anomalous_2023,vanwinkle2025nanoscale}. Experimental measurements have revealed a striking non-monotonic angle-dependence in electrochemical rates, particularly for twist angles below $3^\circ$. An increase in ET rates is observed between $3^\circ$ and 1.5$3^\circ$ as flat bands form with decreasing angle approaching the magic angle. However, a decrease in ET rates is observed with decreasing moiré angle below $1.5^\circ$. This behavior below $1.5^\circ$ is not driven by a continuous evolution of the local electronic environment. Instead, the macroscopic ET rate is primarily a function of the shifting area fraction of the AA-stacking domains. Intriguingly, the local electrochemical activity within these AA domains remains effectively constant, regardless of the global twist angle $\theta_m$. This phenomenon arises from a structural ``pinning" effect: the intense local reconstruction at AA sites forces the atomic geometry in these regions to settle into a nearly invariant configuration, with a local rotation fixed at approximately $1.3–1.6^\circ$regardless of the macroscopic $\theta_m$. From a physical chemistry perspective, this implies that atomic reconstruction creates a universal local chemical environment at the AA topological defect ``active sites" of the moiré electrode; the global twist angle merely acts as a lever to tune the total area density of these active sites across the moiré superlattice. This insight highlights how BI-derived strain maps can be used to understand surface chemical reactivity by quantifying the precisely reconstructed area fractions that govern interfacial charge transfer.

\subsubsection*{Twisted transition metal dichalcogenides}

The application of BI to transition metal dichalcogenide (TMD) superlattices reveals reconstruction mechanics that deviate considerably from those observed in twisted graphene systems.\cite{van2023rotational} A primary distinction arises from the broken inversion symmetry of the constituent TMD monolayers, where parallel (P) and antiparallel (AP) stacking orientations result in fundamentally different moiré symmetries and energy landscapes (Figure 1e,f). While previous atomic-resolution electron microscopy \cite{weston2020atomic} and scanning probe methods \cite{Parkinson1991periodic,rosenberger2020twist,li2021imaging2,sung2022torsional,shabani2021deep,li2021lattice} established the existence of these stacking-dependent features, their characterization was limited to surface topography or maps of interlayer registry. Consequently, the underlying mechanical drivers, specifically the competition between interlayer stacking energy and intralayer elastic deformation, remained largely within the domain of theoretical simulation. \cite{naik2020origin,naik2018ultraflatbands,enaldiev2020stacking,ferreira2021band} BI bridges this gap by providing direct experimental maps of both interlayer displacement fields and local mechanical strain, offering a comprehensive view of the relaxation energetics in both P and AP homobilayers (Figure 5a,b,f,g).\cite{van2023rotational}. 

The structural complexity increases in TMD heterobilayers both lattice mismatch and interlayer twist angle modulate the moiré potential (Figure 1d). In these systems, the moiré landscape is no longer governed by a single parameter. Instead, the interplay among both contributions and strain introduces additional structural degrees of freedom. Because the electronic, optical, and excitonic properties \cite{seyler2019signatures,huang2022excitons,jin2019probing,zhang2020moire,zhang2018moire,tran2019signature,susarla2022hyperspectral,naik2022intralayer,alexeev2019resonantly,wu2018topological,dandu2022electrically} of TMD heterostructures are highly sensitive to local atomic registry, identifying the precise stacking order is essential.

\begin{figure}
\begin{center}
\includegraphics[width=12cm]{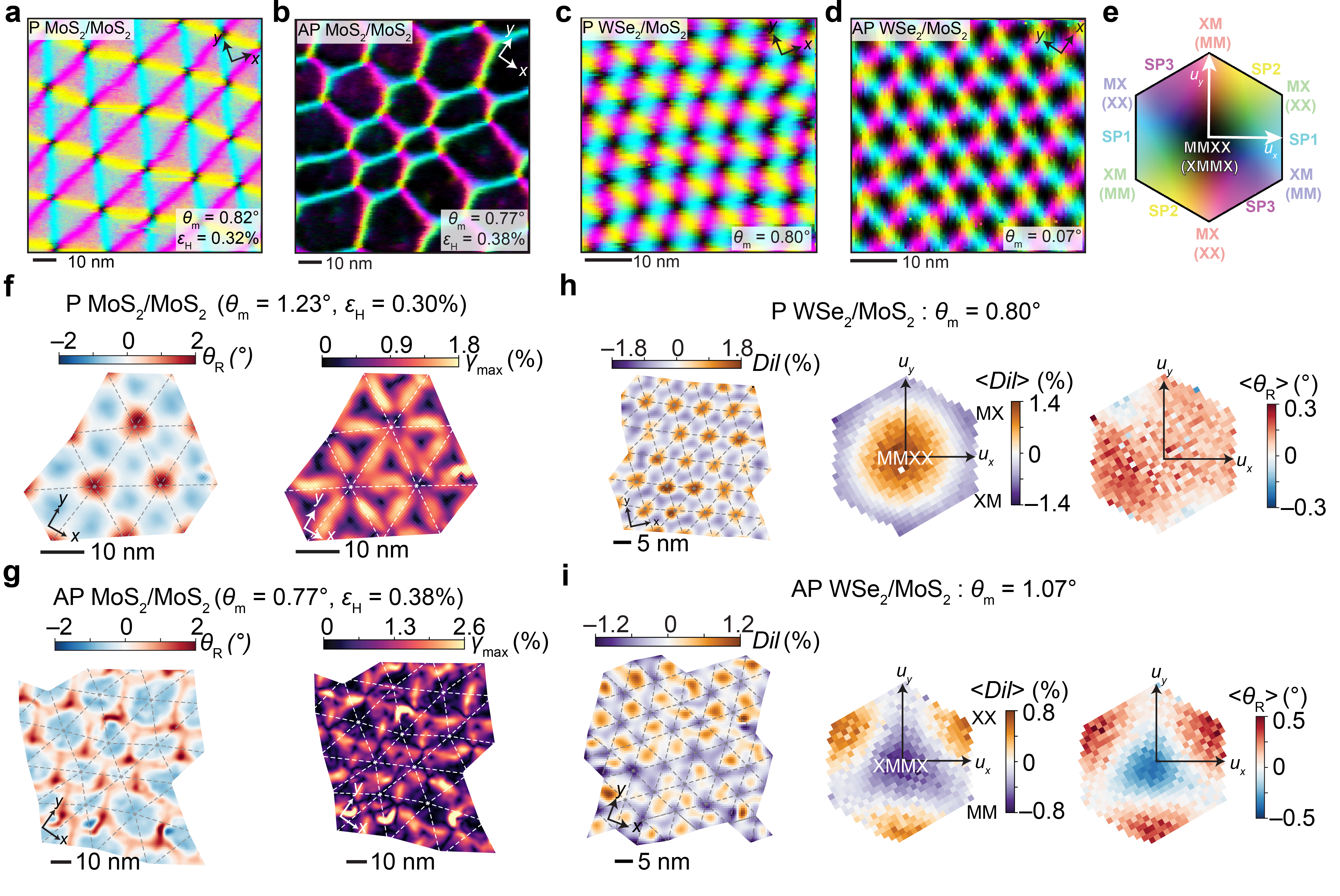}
\end{center}
\caption{  \textbf{Rotational and dilational reconstruction in bilayer TMD moirés.} (a-d) Representative displacement maps for moiré bilayers with parallel (P) and anti-parallel (AP) orientations. The moiré twist angle and heterostrain are labelled as $\theta_m$ and $\varepsilon_H$. (e) 2D displacement hexagon legend for the displacement field maps in a–d. (f, g) Maps of local reconstruction rotation, $\theta_R$ (left), and average principal shear strain, $\gamma_m$ (right) for P and AP MoS$_2$ moiré bilayers. (h, i) Maps of local dilations for P (h) and AP (i) moiré heterobilayers (left). Corresponding 2D plots of average dilation ($<Dil>$) and average reconstruction rotation ($<\theta_R>$) as a function of interlayer displacement $(u_x, u_y)$. Reproduced with permission from reference \cite{van2023rotational}.}
\end{figure}

A key advantage of BI in these materials is that the overlap intensities depend not only on the relative displacement field but also on the crystallographic orientation of the individual layers. In non-centrosymmetric heterobilayers, the coefficients appearing in Equation 10 acquire additional sensitivity to layer orientation through the complex structure factors of the constituent crystals. This allows rotational information that would otherwise be difficult to determine from diffraction intensities alone to be recovered directly from the interference signal. Exploiting this effect, BI determined local rotational reconstruction in twisted TMD heterobilayers and and to separate rotational distortions from other forms of lattice relaxation. Using this capability, the measured displacement fields (Figure 5c,d) revealed that relaxation is not adequately described by purely rotational distortions of the moiré lattice. Instead, substantial dilational and shear components were observed alongside rotational reconstruction (Figure 5h,i), leading to pronounced deviations from the geometry expected from a rigid twist model.\cite{van2023rotational}

The ability of BI to map structural distortions in buried interfaces also unveiled a previously overlooked impact of encapsulating layers. Theoretical models and scanning probe measurements had emphasized out-of-plane stacking-dependent variations in interlayer spacing (out of plane corrugations) as a primary relaxation mechanism. However these studies tended not to use encapsulated moirés, due to the limitations of the experimental probe or computational intractability when encapuslating slabs are included. BI revealed that this out-of-plane pathway is highly sensitive to the mechanical environment. Comparative studies of WSe$_2$/MoS$_2$ heterostructures showed that hBN encapsulation strongly modulates the localization of reconstruction (Figure 6).\cite{van2023rotational} Specifically, fully encapsulated regions (Figure 6a,d,g) exhibit the most robust in-plane rotations and dilations, whereas these deformations are weakest and most diffuse in partially (Figure 6b,e,h) or fully suspended (Figure 6c,f,i) membranes. Encapsulation by hBN layers therefore suppresses out-of-plane corrugation, effectively forcing the system to relax through intensified in-plane pathways. This finding implies that the moiré potential is not intrinsic to the bilayer alone but is fundamentally reshaped by its dielectric and mechanical surroundings, a distinction critical for interpreting the electronic properties of encapsulated quantum devices.

\begin{figure}[h]
\begin{center}
\includegraphics[width=12cm]{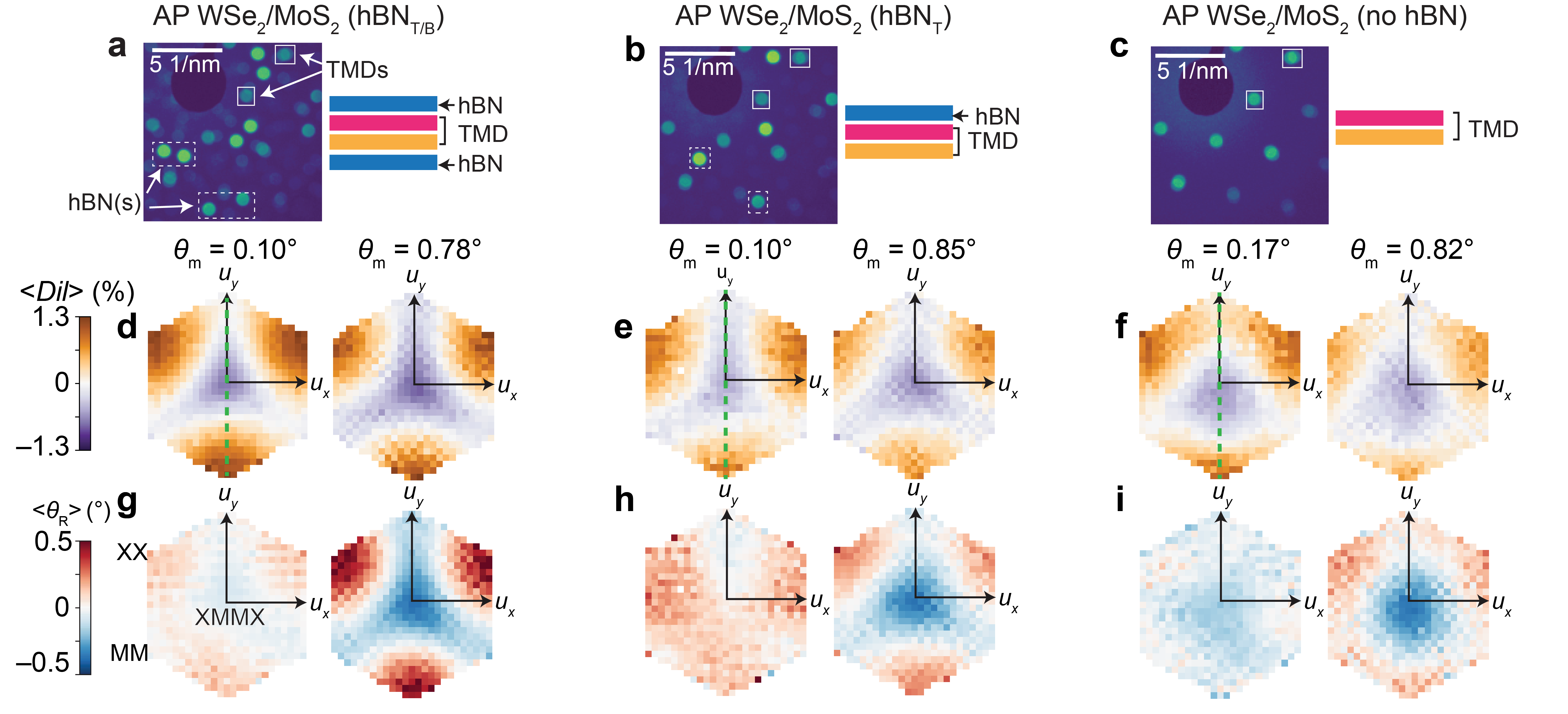}
\end{center}
\caption{  \textbf{Effects of hBN encapsulation on reconstruction.} (a–c) CBED patterns for an AP WSe$_2$/MoS$_2$ sample with three regions: fully encapsulated with hBN (a, $\theta_m = 0.78^\circ$), encapsulated on one side with hBN (b, $\theta_m = 0.85^\circ$), and freely suspended (c, $\theta_m = 0.82^\circ$). (d–f) Average dilations ($<Dil>$) as a function of interlayer displacement $(u_x, u_y)$ for fully capped (d), partially capped (e)), and suspended (f) AP WSe$_2$/MoS$_2$ samples at two twist angles. (g–i) Corresponding 2D plots of plots of average reconstruction rotation ($<\theta_R>$) as a function of displacement. Reproduced with permission from reference \cite{van2023rotational}.} 
\end{figure}

\subsubsection*{Twisted trilayer graphene}

Following the discovery of correlated electron phases in tBG,\cite{cao2018correlated,cao2018unconventional} interest in twisted trilayer graphpene (tTG) increased rapidly because these materials exhibit a broad range of correlated and superconducting phases, often with greater tunability than their bilayer counterparts.\cite{kim2022evidence,park2021tunable,zhu2020twisted,cualuguaru2021twisted,morell2013electronic,turkel2022orderly,phong2021band,uri2023quasicrystal}. What makes tTG particularly complicated is the existence of three moiré wavelengths associated with the three interlayer twist angles (Figure 7a)—in addition to heterostrain—and the increase in local stacking possibilities (Figure 7b). Notwithstanding this structural complexity, experiments had suggested that superconductivity in tTG was more robust to disorder or gating than that found in tBG. Theoretical studies had pointed to a protection of flatbands in tBG \cite{ahn2019failure,song2019all} and tTG \cite{mora2019flatbands, torma2021superfluidity} by space–time inversion ($C_{2z}T$) symmetry, which is present at the AA stacking domains in tBG and AAA domains in tTG. However, these AAA domains should only predominate in tTG samples in which the top and bottom layers are perfectly vertically aligned and mirror symmetric (so-called ``A-twist-A", AtA, samples). However, the observation of superconductivity even in misaligned alternating-twist trilayers \cite{park2021tunable,uri2023quasicrystal} led to the suggestion that misaligned tTG might spontaneously relax to lock into a mirror symmetric structure,\cite{turkel2022orderly} motivating a direct probe of interlayer stacking after relaxation.

\begin{figure}
\begin{center}
\includegraphics[width=11cm]{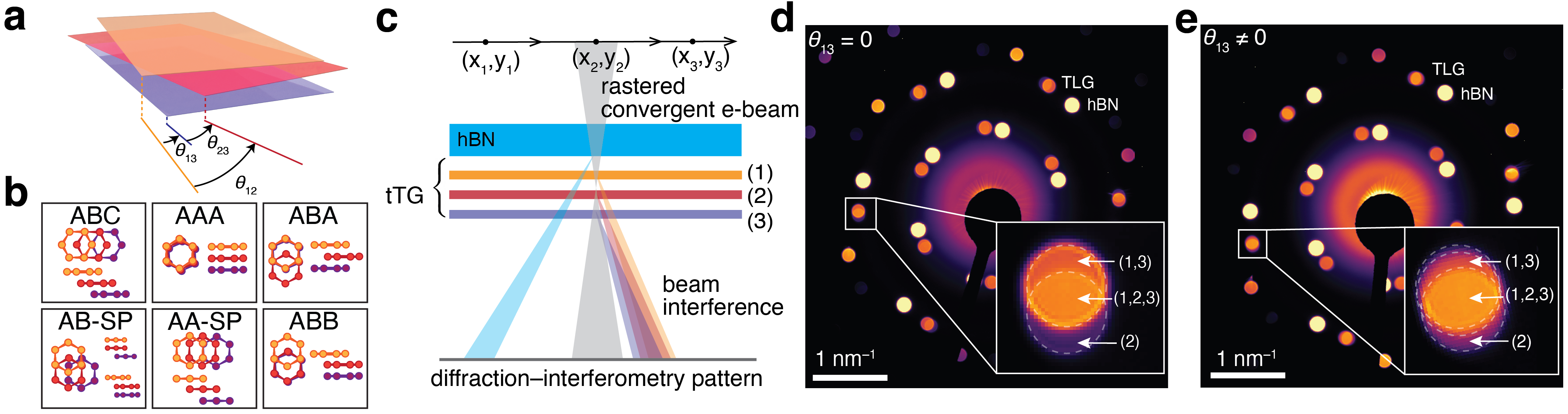}
\end{center}
\caption{  \textbf{TTG stacking and CBED patterns.} (a) Schematic illustrating the twist angle, $\theta$, and layer numbering conventions used to label the graphene trilayers. (b) Illustrations of various high-symmetry stacking configurations realized within tTG. (c) Schematic of 4D-STEM measurement in a tTG sample (d,e) Average CBED patterns for tTG with $\theta_{13} \approx 0^\circ$ (d) and $\theta_{13} \approx 0.22^\circ$ (e). Overlapping tTG Bragg disks are highlighted in the insets. Reproduced with permission from reference \cite{craig2024local}.} 
\end{figure}

The application of Bragg interferometry (Figure 7c) to tTG demonstrates the versatility of the method in disentangling multi-layer stacking geometries.\cite{craig2024local} In a trilayer system, each constituent layer generates a unique set of Bragg disks, resulting in a complex diffraction pattern characterized by multiple, distinct overlap regions (Figure 7d,e). By selectively analyzing the interference intensities within specific overlap zones—such as those involving only layers 1 and 3 (insets of Figure 7d,e)—BI can, in principle, treat these regions as isolated bilayer interferograms. This makes BI a powerful tool for directly tackling the problem of relaxation misaligned trilayers. 

\begin{figure}
\begin{center}
\includegraphics[width=11cm]{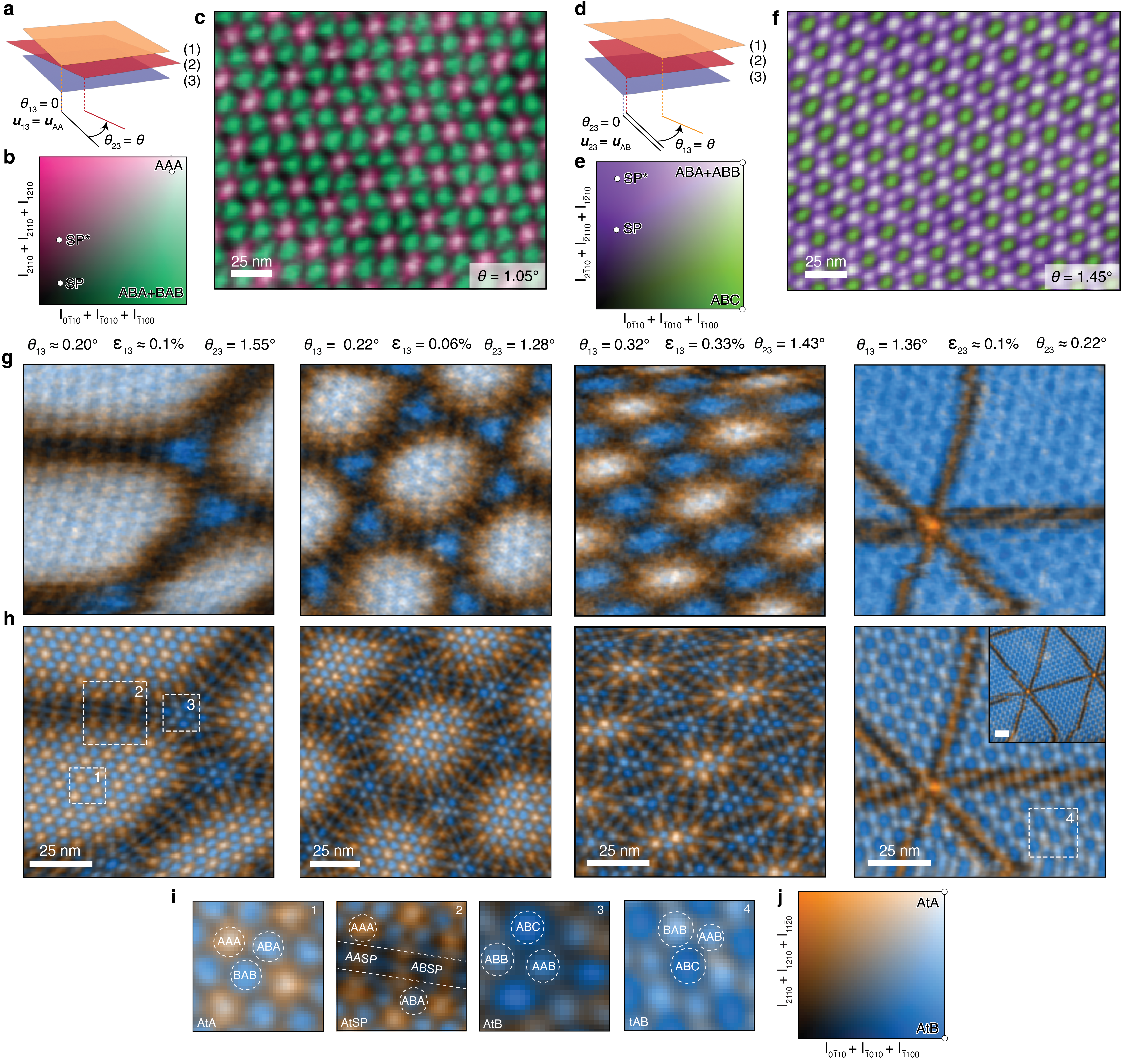}
\end{center}
\caption{  \textbf{Imaging reconstruction in tTG.} (a,d) Layer stacking schematics for two limiting configurations of tTG. (b,e,j) Legends illustrating how colour correlates with the average
first- and second-order Bragg disc intensities. Overlaid points are the intensities of high-symmetry stacking orders. (c,f) Maps of local stacking order in samples with limiting configurations of tTG. (g) Maps of local atomic stacking from the larger moiré pattern only, corresponding to the local in-plane offset between layers 1 and 3 in columns 1-3, and the local in-plane offset between layers 2 and 3 in column 4. Colours shown correspond to the bivariate colour map in j. (h) Local atomic stacking, over the same regions as those shown in g, obtained from considering all three graphene layers; the inset at the far right shows a zoomed-out view of the structure in column 4 of g. (i) Zoomed-in views of the numbered dashed boxes in the maps of b above illustrating the finer local modulation of stacking order. Reproduced with permission from reference \cite{craig2024local}.} 
\end{figure}

In trilayer architectures where two of the three layers are crystallographically aligned—such as the high-symmetry AtA (Figure 8a) and tAB (Figure 8d) configurations—BI maps of the local stacking registry reveal a singular moiré periodicity (Figure 8c,f). While this periodicity mirrors the predictions of rigid-lattice models, the experimental stacking area fractions deviate significantly due to atomic relaxation. Specifically, BI reveals a pronounced contraction of high-energy stacking domains: the AAA sites in AtA samples and the AAB sites in tAB samples are significantly reduced in size. These observations provide direct experimental evidence of the energetic favorability of local ABA/BAB-type registry in AtA samples and that of ABC/BAB domains in tAB samples, demonstrating how the system minimizes its global energy by expanding stable stacking regions at the expense of unfavorable, vertically aligned configurations.\cite{craig2024local} 

In tTG samples that deviate from perfectly aligned end-member polytypes, the structural landscape becomes significantly more complex. By leveraging the double-overlap regions within the CBED patterns to map the local stacking registry, BI can isolate the relative alignment responsible for long- (Figure 8g) and short- (Figure 8h) period moiré patterns. These maps reveal that while atomic relaxation in misaligned, alternating-twist trilayers increases the area fraction of mirror-symmetric AtA domains (white regions), it does not entirely eliminate the AtB stacking motifs (blue regions). This observation stands in contrast to earlier STM findings, suggesting a more complete transition toward AtA order \cite{turkel2022orderly}.

Furthermore, the measured preference for AtA-type stacking over AtB-type and soliton-like (grey) regions was unexpected. In rigid-lattice models, the energetic difference between AtA and AtB stacking—determined solely by the weak coupling between the non-adjacent top and bottom layers—was previously presumed to be a negligible driver of reconstruction \cite{zhu2020twisted}. The discovery that this preference persists experimentally demonstrates that trilayer reconstruction is not merely an additive extension of bilayer mechanics. Instead, the displacement fields across all three layers are fundamentally coupled, indicating that even second-neighbor interlayer interactions play a non-trivial role in defining the global energy minimum of the superlattice.

This trilayer case also highlights a broader methodological strength of BI. Because the measured interference signal contains contributions from each pair of layers, the approach naturally generalizes to multilayer structures without requiring atomic-resolution imaging of every interface. This feature makes the formalism readily extendable to more complex multilayer moiré heterostructures and buried interfaces.

\subsection*{Limitations and Implementation Challenges}

Despite the successful application of Bragg Interferometry to a growing range of moiré systems, several challenges remain before it (and related dark-field approaches) can be considered a routine tool for quantitative structural chacterization. Many of these constraints arise from the very ``coarse-graining" that makes the technique computationally efficient: by fitting only a subset of the total scattering information, the method necessitates certain physical assumptions.

\subsubsection*{The Phase and Reference Problem}
A primary challenge is the incomplete phase information available from dark-field measurements. As discussed previously, BI determines relative geometric phases between layers through the interference of overlapping Bragg reflections. Unlike GPA or full ptychographic reconstruction, however, the absolute phase information contained within the bright-field disk is not directly measured. Consequently, interpretation of the measured displacement fields requires assumptions regarding the average lattice constant, reference structure, and partitioning of deformation between layers. These assumptions are often reasonable for weakly strained bilayers but become increasingly difficult to justify in strongly reconstructed or multilayer systems.

\subsubsection*{Dynamical Scattering and Sample Thickness}
Another important limitation concerns the weak phase object approximation used in most current implementations, assuming that multiple scattering is negligible and that the measured intensities can be interpreted through interference between independently scattered waves. While this approximation is well satisfied for many atomically thin materials, increasing sample thickness, heavier constituent elements, or operation near strong diffraction conditions can introduce dynamical scattering effects that modify the overlap intensities. Future iterations of the methodology will benefit from the integration of multislice simulations directly into the fitting routine to account for these non-linear scattering contributions.

\subsubsection*{Geometric Projections and Tilt}
The role of specimen tilt and three-dimensional geometry represents a significant confounding factor. Because moiré materials are atomically thin, their diffraction peaks are elongated into reciprocal lattice rods (so-called ``relrods"). Small local tilts alter the intersection of these relrods with the Ewald sphere, causing streaks and intensity shifts that can be indistinguishable from genuine in-plane strain. Furthermore, the measured displacement $\mathbf{u}$ is merely an in-plane projection; without a rigorous treatment of sample orientation, local tilts can lead to an incorrect mapping of the true relaxation.

\subsection*{Future directions}
Beyond these limitations, there are also substantial opportunities for methodological development. The evolution of moiré characterization lies in moving beyond the rigid bilayer paradigm toward a holistic, three-dimensional understanding of superlattice mechanics.

\subsubsection*{Simultaneous In-plane and Out-of-plane Mapping}
Vertical relaxation (out-of-plane corrugation and variations in interlayer spacing) can play as important a role in electronic structure as in-plane strain, as shown by the fact that hBN encapsulation modifies in-plane TMD moiré relaxation. As discussed in the context of CBED holography, out-of-plane separation modulates the phase relationships between interfering beams and therefore contributes to the observed fringe structure. A key future direction is the development of BI such that the method is capable of simultaneously measuring $u_z$ (corrugation) along with $u_x$ and $u_y$ (strain), providing a complete picture of the 3D reconstruction and potential landscape.

\subsubsection*{Complexity in Multilayer Heterostructures}
As the field shifts toward increasing numbers of layers,\cite{park2022robust,fujimoto2025four,Han2024correlated} an understanding of reconstruction in more complex multilayer systems will be essential. While recent extensions discussed here have demonstrated that BI can be generalized beyond bilayers, the number of possible pairwise displacement fields grows combinatorially with increasing number of layers. This complexity introduces ambiguities that may require additional constraints from symmetry, elasticity theory, or complementary experimental measurements. Developing robust reconstruction procedures for multilayer heterostructures will likely become increasingly important as the sophistication of experimentally realized moiré systems continues to grow.

\subsubsection*{Hybrid Methodologies}
There is substantial opportunity in the integration of BI with complementary phase-sensitive techniques and other structural and electronic probes. GPA, CBED holography, and ptychographic approaches each provide access to different portions of the available structural information and rely on different assumptions regarding phase recovery. Combining these approaches within a common framework may enable more complete characterization than any individual method alone. For example, bright-field measurements could provide information regarding absolute lattice distortions while dark-field interferometric measurements retain sensitivity to buried interfaces and encapsulated structures. Combining BI with magnetic imaging (e.g., magnetic force microscopy, spin-polarized STM, and lorentz TEM) as well as near-field optical or electron-based spectroscopic measurements will also shed important light on how local structure impacts exotic physical behavior.

\subsubsection*{Machine Learning and Artificial Intelligence}
As moiré systems increase in complexity, moving toward thicker multi-layer heterostructures and non-linear dynamical scattering regimes, the integration of artificial intelligence (AI) offers a path to bypass the traditional reliance on manual analytical fitting and the weak phase object approximation. Deep learning architectures, particularly Physics-Informed Neural Networks, could be useful for phase unwrapping of complex, noisy displacement fields by enforcing the aforementioned physical constraints (elasticity theory or symmetry requirements) directly into the optimization process. Furthermore, convolutional neural networks can be trained on vast libraries of 4D-STEM datasets and multislice simulations to recognize the subtle signatures of dynamical scattering or out-of-plane corrugation that are currently intractable by analytical models. Such AI-driven workflows could help automate the identification of stacking motifs and the quantification of strain across micron-scale fields of view, essentially acting as an intelligent inverse solver for the 4D-STEM dataset. Ultimately, shifting toward AI-assisted reconstruction may prove essential for transforming Bragg Interferometry from a specialized tool into a high-throughput standard for characterizing the intricate structural foundations of quantum materials (even beyond moiré materials to other materials with co-localized reciprocal lattice vectors) and its integration with \textit{operando} measurements.

\section*{Conclusions}
The characterization of moiré superlattices represents a modern frontier in materials science, chemical physics, and condensed matter physics, where the macroscopic electronic, optical, magnetic, and (electro)chemical properties of a device are dictated by sub-angstrom atomic relaxations at buried interfaces. In this review, we have contextualized the development of Bragg interferometry within a broader lineage of electron microscopy and X-ray spectroscopy. 

By comparing a family of phase-retrieval methods (from full ptychographic reconstruction of the specimen's complex transmission function to targeted extraction of geometric phase information in HRTEM images and overlapping Bragg reflections) we have highlighted the advantages of a constrained approach that is particularly well suited to the unique hierarchy of length scales in 2D moiré heterostructures, where nanoscale structural modulations propagate, usually non-uniformly, over micron-scale distances.

The application of BI to moiré graphene and TMD systems has already provided pivotal insights that challenge rigid-lattice assumptions. These include the discovery of reconstruction-stabilized flat bands in magic-angle tBG, the observation of dilational reconstruction and environment-dependent relaxation pathways in encapsulated TMD heterobilayers, and persistent second-neighbor interlayer interactions in tTG. However, these case studies also underscore the inherent complexities of relying exclusively on dark-field information. While this approach offers a decisive advantage in isolating the signal of buried interfaces from strongly scattering ancillary layers or substrates, it necessitates a rigorous handling of incomplete phase information, dynamical scattering, and three-dimensional geometric projections.

Looking forward, the maturation of BI will likely coincide with increasingly sophisticated ``twistronic" architectures. As structural degrees of freedom in these materials multiply, the field must transition toward hybrid methodologies that integrate interferometric precision and versatility with other probes and global optimization capabilities of AI. By bridging the gap between fundamental diffraction physics and large-scale structural mapping, Bragg interferometry can serve as a cornerstone technique, transforming our ability to visualize and engineer the structural foundations of emergent quantum phenomena.

\section*{Disclosure statement}
The authors are not aware of any affiliations, memberships, funding, or financial holdings that
might be perceived as affecting the objectivity of this review. 

\section*{Acknowledgments}
We acknowledge helpful conversations and formative contributions from M. Van Winkle, N. Kazimerzak, C. Ophus, K. Bustillo, J. Ciston, H. Brown, C. Groshner, T. Taniguchi, and K. Watanabe in the referenced initial realizations of this approach. This material is based upon work supported by the US National Science Foundation Early Career Development Program (CAREER), under award no. 2238196. 

\section*{Literature\ cited}

\printbibliography

@article{suarez2010flat,
  title = {Flat bands in slightly twisted bilayer graphene: Tight-binding calculations},
  author = {Su\'arez Morell, E. and Correa, J. D. and Vargas, P. and Pacheco, M. and Barticevic, Z.},
  journal = {Phys. Rev. B},
  volume = {82},
  pages = {121407(R)},
  year = {2010},
  publisher = {American Physical Society},

}

@article{Parkinson1991periodic,
    author = {Parkinson, B. A. and Ohuchi, F. S. and Ueno, K. and Koma, A.},
    title = {Periodic lattice distortions as a result of lattice mismatch in epitaxial films of two‐dimensional materials},
    journal = {Applied Physics Letters},
    volume = {58},
    pages = {472-474},
    year = {1991},
}

@article{nakatsuji2023multi,
  title={Multi-scale lattice relaxation in chiral twisted trilayer graphenes},
  author={Nakatsuji, Naoto and Kawakami, Takuto and Koshino, Mikito},
  journal={arXiv preprint arXiv:2305.13155},
  year={2023}
}

@article{chen2019signatures,
  title={Signatures of tunable superconductivity in a trilayer graphene moir{\'e} superlattice},
  author={Chen, Guorui and Sharpe, Aaron L and Gallagher, Patrick and Rosen, Ilan T and Fox, Eli J and Jiang, Lili and Lyu, Bosai and Li, Hongyuan and Watanabe, Kenji and Taniguchi, Takashi and others},
  journal={Nature},
  volume={572},
  number={7768},
  pages={215--219},
  year={2019},
  publisher={Nature Publishing Group UK London}
}

@article{park2021tunable,
  title={Tunable strongly coupled superconductivity in magic-angle twisted trilayer graphene},
  author={Park, Jeong Min and Cao, Yuan and Watanabe, Kenji and Taniguchi, Takashi and Jarillo-Herrero, Pablo},
  journal={Nature},
  volume={590},
  number={7845},
  pages={249--255},
  year={2021},
  publisher={Nature Publishing Group UK London}
}

@article{cao2018correlated,
  title={Correlated insulator behaviour at half-filling in magic-angle graphene superlattices},
  author={Cao, Yuan and Fatemi, Valla and Demir, Ahmet and Fang, Shiang and Tomarken, Spencer L and Luo, Jason Y and Sanchez-Yamagishi, Javier D and Watanabe, Kenji and Taniguchi, Takashi and Kaxiras, Efthimios and others},
  journal={Nature},
  volume={556},
  number={7699},
  pages={80--84},
  year={2018},
  publisher={Nature Publishing Group}
}

@article{cao2018unconventional,
	author = {Cao, Yuan and Fatemi, Valla and Fang, Shiang and Watanabe, Kenji and Taniguchi, Takashi and Kaxiras, Efthimios and Jarillo-Herrero, Pablo},
	journal = {Nature},
	number = {7699},
	pages = {43--50},
	title = {Unconventional superconductivity in magic-angle graphene superlattices},
	volume = {556},
	year = {2018}
 
 }

@article{morell2013electronic,
  title={Electronic properties of twisted trilayer graphene},
  author={Morell, E Su{\'a}rez and Pacheco, M and Chico, Leonor and Brey, Luis},
  journal={Physical Review B},
  volume={87},
  number={12},
  pages={125414},
  year={2013},
  publisher={APS}
}

@article{hao2021electric,
  title={Electric field--tunable superconductivity in alternating-twist magic-angle trilayer graphene},
  author={Hao, Zeyu and Zimmerman, AM and Ledwith, Patrick and Khalaf, Eslam and Najafabadi, Danial Haie and Watanabe, Kenji and Taniguchi, Takashi and Vishwanath, Ashvin and Kim, Philip},
  journal={Science},
  volume={371},
  number={6534},
  pages={1133--1138},
  year={2021},
  publisher={American Association for the Advancement of Science}
}

@article{kim2022evidence,
  title={Evidence for unconventional superconductivity in twisted trilayer graphene},
  author={Kim, Hyunjin and Choi, Youngjoon and Lewandowski, Cyprian and Thomson, Alex and Zhang, Yiran and Polski, Robert and Watanabe, Kenji and Taniguchi, Takashi and Alicea, Jason and Nadj-Perge, Stevan},
  journal={Nature},
  volume={606},
  number={7914},
  pages={494--500},
  year={2022},
  publisher={Nature Publishing Group UK London}
}

@article{zhu2020twisted,
  title={Twisted trilayer graphene: A precisely tunable platform for correlated electrons},
  author={Zhu, Ziyan and Carr, Stephen and Massatt, Daniel and Luskin, Mitchell and Kaxiras, Efthimios},
  journal={Physical review letters},
  volume={125},
  number={11},
  pages={116404},
  year={2020},
  publisher={APS}
}

@article{kazmierczak2021strain,
  title={Strain fields in twisted bilayer graphene},
  author={Kazmierczak, Nathanael P and Van Winkle, Madeline and Ophus, Colin and Bustillo, Karen C and Carr, Stephen and Brown, Hamish G and Ciston, Jim and Taniguchi, Takashi and Watanabe, Kenji and Bediako, D Kwabena},
  journal={Nature materials},
  volume={20},
  number={7},
  pages={956--963},
  year={2021},
  publisher={Nature Publishing Group UK London}
}

@article{turkel2022orderly,
  title={Orderly disorder in magic-angle twisted trilayer graphene},
  author={Turkel, Simon and Swann, Joshua and Zhu, Ziyan and Christos, Maine and Watanabe, K and Taniguchi, T and Sachdev, Subir and Scheurer, Mathias S and Kaxiras, Efthimios and Dean, Cory R and others},
  journal={Science},
  volume={376},
  number={6589},
  pages={193--199},
  year={2022},
  publisher={American Association for the Advancement of Science}
}

@article{balents2020superconductivity,
  title={Superconductivity and strong correlations in moir{\'e} flat bands},
  author={Balents, Leon and Dean, Cory R and Efetov, Dmitri K and Young, Andrea F},
  journal={Nature Physics},
  volume={16},
  number={7},
  pages={725--733},
  year={2020},
  publisher={Nature Publishing Group UK London}
}

@article{lau2022reproducibility,
  title={Reproducibility in the fabrication and physics of moir{\'e} materials},
  author={Lau, Chun Ning and Bockrath, Marc W and Mak, Kin Fai and Zhang, Fan},
  journal={Nature},
  volume={602},
  number={7895},
  pages={41--50},
  year={2022},
  publisher={Nature Publishing Group UK London}
}

@article{zachman2021interferometric,
  title={Interferometric 4D-STEM for lattice distortion and interlayer spacing measurements of bilayer and trilayer 2D materials},
  author={Zachman, Michael J and Madsen, Jacob and Zhang, Xiang and Ajayan, Pulickel M and Susi, Toma and Chi, Miaofang},
  journal={Small},
  volume={17},
  number={28},
  pages={2100388},
  year={2021},
  publisher={Wiley Online Library}
}

@article{song2019all,
  title={All magic angles in twisted bilayer graphene are topological},
  author={Song, Zhida and Wang, Zhijun and Shi, Wujun and Li, Gang and Fang, Chen and Bernevig, B Andrei},
  journal={Physical review letters},
  volume={123},
  number={3},
  pages={036401},
  year={2019},
  publisher={APS}
}

@article{ahn2019failure,
  title={Failure of Nielsen-Ninomiya theorem and fragile topology in two-dimensional systems with space-time inversion symmetry: application to twisted bilayer graphene at magic angle},
  author={Ahn, Junyeong and Park, Sungjoon and Yang, Bohm-Jung},
  journal={Physical Review X},
  volume={9},
  number={2},
  pages={021013},
  year={2019},
  publisher={APS}
}

@article{mora2019flatbands,
  title={Flatbands and perfect metal in trilayer moir{\'e} graphene},
  author={Mora, Christophe and Regnault, Nicolas and Bernevig, B Andrei},
  journal={Physical review letters},
  volume={123},
  number={2},
  pages={026402},
  year={2019},
  publisher={APS}
}

@article{cualuguaru2021twisted,
  title={Twisted symmetric trilayer graphene: Single-particle and many-body Hamiltonians and hidden nonlocal symmetries of trilayer moir{\'e} systems with and without displacement field},
  author={C{\u{a}}lug{\u{a}}ru, Dumitru and Xie, Fang and Song, Zhi-Da and Lian, Biao and Regnault, Nicolas and Bernevig, B Andrei},
  journal={Physical Review B},
  volume={103},
  number={19},
  pages={195411},
  year={2021},
  publisher={APS}
}

@article{phong2021band,
  title={Band structure and superconductivity in twisted trilayer graphene},
  author={Phong, V{\~o} Tiến and Pantale{\'o}n, Pierre A and Cea, Tommaso and Guinea, Francisco},
  journal={Physical Review B},
  volume={104},
  number={12},
  pages={L121116},
  year={2021},
  publisher={APS}
}

@article{kerelsky2019maximized,
  title={Maximized electron interactions at the magic angle in twisted bilayer graphene},
  author={Kerelsky, Alexander and McGilly, Leo J and Kennes, Dante M and Xian, Lede and Yankowitz, Matthew and Chen, Shaowen and Watanabe, K and Taniguchi, T and Hone, James and Dean, Cory and others},
  journal={Nature},
  volume={572},
  number={7767},
  pages={95--100},
  year={2019},
  publisher={Nature Publishing Group UK London}
}

@article{carr2018relaxation,
  title = {Relaxation and domain formation in incommensurate two-dimensional heterostructures},
  author = {Carr, Stephen and Massatt, Daniel and Torrisi, Steven B. and Cazeaux, Paul and Luskin, Mitchell and Kaxiras, Efthimios},
  journal = {Phys. Rev. B},
  volume = {98},
  issue = {22},
  pages = {224102},
  numpages = {7},
  year = {2018},
  publisher = {American Physical Society},
}

@article{torma2021superfluidity,
	Author = {T{\"o}rm{\"a}, P{\"a}ivi and Peotta, Sebastiano and Bernevig, Bogdan A.},
	Journal = {Nature Reviews Physics},
	Pages = {528--542},
	Title = {Superconductivity, superfluidity and quantum geometry in twisted multilayer systems},
	Volume = {4},
	Year = {2022}}

@article{yoo2019atomic,
  title={Atomic and electronic reconstruction at the van der Waals interface in twisted bilayer graphene},
  author={Yoo, Hyobin and Engelke, Rebecca and Carr, Stephen and Fang, Shiang and Zhang, Kuan and Cazeaux, Paul and Sung, Suk Hyun and Hovden, Robert and Tsen, Adam W and Taniguchi, Takashi and others},
  journal={Nat. Mater.},
  volume={18},
  number={5},
  pages={448--453},
  year={2019},
  publisher={Nature Publishing Group}
}

@article{uri2023quasicrystal,
	Author = {Uri, Aviram and de la Barrera, Sergio C. and Randeria, Mallika T. and Rodan-Legrain, Daniel and Devakul, Trithep and Crowley, Philip J. D. and Paul, Nisarga and Watanabe, Kenji and Taniguchi, Takashi and Lifshitz, Ron and Fu, Liang and Ashoori, Raymond C. and Jarillo-Herrero, Pablo},
	Journal = {Nature},
	Number = {7975},
	Pages = {762--767},
	Title = {Superconductivity and strong interactions in a tunable moir{\'e}quasicrystal},
	Volume = {620},
	Year = {2023}
}

@article{ophus2019four,
  title={Four-dimensional scanning transmission electron microscopy (4D-STEM): From scanning nanodiffraction to ptychography and beyond},
  author={Ophus, Colin},
  journal={Microscopy and Microanalysis},
  volume={25},
  number={3},
  pages={563--582},
  year={2019},
  publisher={Oxford University Press}
}

@article{huang2022excitons,
  author  = {Huang, D. and Choi, J. and Shih, C.-K. and Li, X.},
  title   = {Excitons in semiconductor moir{\'e} superlattices},
  journal = {Nature Nanotechnology},
  volume  = {17},
  pages   = {227--238},
  year    = {2022}
}

@article{mak2022semiconductor,
  author  = {Mak, K.~F. and Shan, J.},
  title   = {Semiconductor moir{\'e} materials},
  journal = {Nature Nanotechnology},
  volume  = {17},
  pages   = {686--696},
  year    = {2022}
}

@article{zhang2018moire,
  author  = {Zhang, N. and Surrente, A. and Baranowski, M. and Maude, D.~K. and Gant, P. and Castellanos-Gomez, A. and Plochocka, P.},
  title   = {Moir{\'e} intralayer excitons in a $\mathrm{MoSe_2}$/$\mathrm{MoS_2}$ heterostructure},
  journal = {Nano Letters},
  volume  = {18},
  pages   = {7651--7657},
  year    = {2018}
}

@article{seyler2019signatures,
  author  = {Seyler, K.~L. and Rivera, P. and Yu, H. and Wilson, N.~P. and Ray, E.~L. and Mandrus, D.~G. and Yan, J. and Yao, W. and Xu, X.},
  title   = {Signatures of moir{\'e}-trapped valley excitons in $\mathrm{MoSe_2}$/$\mathrm{WSe_2}$ heterobilayers},
  journal = {Nature},
  volume  = {567},
  pages   = {66--70},
  year    = {2019}
}

@article{alexeev2019resonantly,
  author  = {Alexeev, E.~M. and Ruiz-Tijerina, D.~A. and Danovich, M. and Hamer, M.~J. and Terry, D.~J. and Nayak, P.~K. and Ahn, S. and Pak, S. and Lee, J. and Sohn, J.~I. and others},
  title   = {Resonantly hybridized excitons in moir{\'e} superlattices in van der Waals heterostructures},
  journal = {Nature},
  volume  = {567},
  pages   = {81--86},
  year    = {2019}
}

@article{jin2019observation,
  author  = {Jin, C. and Regan, E.~C. and Yan, A. and Iqbal Bakti~Utama, M. and Wang, D. and Zhao, S. and Qin, Y. and Yang, S. and Zheng, Z. and Shi, S. and others},
  title   = {Observation of moir{\'e} excitons in $\mathrm{WSe_2}$/$\mathrm{WS_2}$ heterostructure superlattices},
  journal = {Nature},
  volume  = {567},
  pages   = {76--80},
  year    = {2019}
}

@article{tran2019evidence,
  author  = {Tran, K. and Moody, G. and Wu, F. and Lu, X. and Choi, J. and Kim, K. and Rai, A. and Sanchez, D.~A. and Quan, J. and Singh, A. and others},
  title   = {Evidence for moir{\'e} excitons in van der Waals heterostructures},
  journal = {Nature},
  volume  = {567},
  pages   = {71--75},
  year    = {2019}
}

@article{dandu2022electrically,
  author = {Dandu, M. and Gupta, G. and Dasika, P. and Watanabe, K. and Taniguchi, T. and Majumdar, K.},
  title = {Electrically tunable localized versus delocalized intralayer moir{\'e} excitons and trions in a twisted MoS$_2$ bilayer},
  journal = {ACS Nano},
  volume = {16},
  pages = {8983--8992},
  year = {2022}
}

@article{naik2022intralayer,
  author = {Naik, M. H. and Regan, E. C. and Zhang, Z. and Chan, Y.-H. and Li, Z. and Wang, D. and Yoon, Y. and Ong, C. S. and Zhao, W. and Zhao, S. and others},
  title = {Intralayer charge-transfer moir{\'e} excitons in van der Waals superlattices},
  journal = {Nature},
  volume = {609},
  pages = {52--57},
  year = {2022}
}

@article{susarla2022hyperspectral,
	Author = {Susarla, Sandhya and Naik, Mit H. and Blach, Daria D. and Zipfel, Jonas and Taniguchi, Takashi and Watanabe, Kenji and Huang, Libai and Ramesh, Ramamoorthy and da Jornada, Felipe H. and Louie, Steven G. and Ercius, Peter and Raja, Archana},
	Journal = {Science},
	Pages = {1235--1239},
	Title = {Hyperspectral imaging of exciton confinement within a moir{\'e} unit cell with a subnanometer electron probe},
	Volume = {378},
	Year = {2022}}

@article{tang2020simulation,
  author = {Tang, Y. and Li, L. and Li, T. and Xu, Y. and Liu, S. and Barmak, K. and Watanabe, K. and Taniguchi, T. and MacDonald, A. H. and Shan, J. and others},
  title = {Simulation of Hubbard model physics in WSe$_2$/WS$_2$ moir{\'e} superlattices},
  journal = {Nature},
  volume = {579},
  pages = {353--358},
  year = {2020}
}

@article{kang2024evidence,
  author = {Kang, K. and Shen, B. and Qiu, Y. and Zeng, Y. and Xia, Z. and Watanabe, K. and Taniguchi, T. and Shan, J. and Mak, K. F.},
  title = {Evidence of the fractional quantum spin Hall effect in moir{\'e} MoTe$_2$},
  journal = {Nature},
  volume = {628},
  pages = {522--526},
  year = {2024}
}

@article{wu2018topological,
  author = {Wu and Lovorn and MacDonald},
  title = {Topological exciton bands in moiré heterojunctions},
  journal = {Physical Review Letters},
  volume = {118},
  pages = {147401},
  year = {2017}
}

@article{zhang2020moire,
  author = {Zhang and Mao and Cao and others},
  title = {Moiré excitons in van der Waals heterostructures},
  journal = {Nature Communications},
  volume = {11},
  pages = {5888},
  year = {2020}
}

@article{bistritzer2011moire,
  author = {Bistritzer and MacDonald},
  title = {Moiré bands in twisted double-layer graphene},
  journal = {Proceedings of the National Academy of Sciences},
  volume = {108},
  pages = {12233--12237},
  year = {2011}
}

@article{jin2019probing,
  author = {Jin and Regan and Yan and Iqbal Bakti Utama and Wang and Zhao and Qin and Yang and Zheng and Shi and others},
  title = {Probing moiré excitons in van der Waals heterostructures},
  journal = {Nature},
  volume = {567},
  pages = {76--80},
  year = {2019}
}

@article{tran2019signature,
  author = {Tran and Moody and Wu and Lu and Choi and Kim and Rai and Sanchez and Quan and Singh and others},
  title = {Signature of moiré excitons in van der Waals heterostructures},
  journal = {Nature},
  volume = {567},
  pages = {71--75},
  year = {2019}
}

@article{weston2020atomic,
  title={Atomic reconstruction in twisted bilayers of transition metal dichalcogenides},
  author={Weston, Astrid and Zou, Yichao and Enaldiev, Vladimir and Summerfield, Alex and Clark, Nicholas and Z{\'o}lyomi, Viktor and Graham, Abigail and Yelgel, Celal and Magorrian, Samuel and Zhou, Mingwei and others},
  journal={Nat. Nanotechnol.},
  volume={15},
  number={7},
  pages={592--597},
  year={2020},
  publisher={Nature Publishing Group}
}

@article{rosenberger2020twist,
  title={Twist angle-dependent atomic reconstruction and moir{\'e} patterns in transition metal dichalcogenide heterostructures},
  author={Rosenberger, Matthew R and Chuang, Hsun-Jen and Phillips, Madeleine and Oleshko, Vladimir P and McCreary, Kathleen M and Sivaram, Saujan V and Hellberg, C Stephen and Jonker, Berend T},
  journal={ACS Nano},
  volume={14},
  number={4},
  pages={4550--4558},
  year={2020},
  publisher={ACS Publications}
}

@article{naik2020origin,
  title={Origin and evolution of ultraflat bands in twisted bilayer transition metal dichalcogenides: Realization of triangular quantum dots},
  author={Naik, Mit H and Kundu, Sudipta and Maity, Indrajit and Jain, Manish},
  journal={Phys. Rev. B},
  volume={102},
  number={7},
  pages={075413},
  year={2020},
  publisher={APS}
}

@article{naik2018ultraflatbands,
  title={Ultraflatbands and shear solitons in moir{\'e} patterns of twisted bilayer transition metal dichalcogenides},
  author={Naik, Mit H and Jain, Manish},
  journal={Phys. Rev. Lett.},
  volume={121},
  number={26},
  pages={266401},
  year={2018},
  publisher={APS}
}

@article{enaldiev2020stacking,
  title={Stacking domains and dislocation networks in marginally twisted bilayers of transition metal dichalcogenides},
  author={Enaldiev, VV and Z{\'o}lyomi, Viktor and Yelgel, CELAL and Magorrian, SJ and Fal’ko, VI},
  journal={Phys. Rev. Lett.},
  volume={124},
  number={20},
  pages={206101},
  year={2020},
  publisher={APS}
}

@article{ferreira2021band,
  title={Band energy landscapes in twisted homobilayers of transition metal dichalcogenides},
  author={Ferreira, F{\'a}bio and Magorrian, SJ and Enaldiev, VV and Ruiz-Tijerina, DA and Fal'ko, VI},
  journal={Appl. Phys. Lett.},
  volume={118},
  number={24},
  pages={241602},
  year={2021},
  publisher={AIP Publishing LLC}
}

@article{li2021lattice,
  title={Lattice reconstruction induced multiple ultra-flat bands in twisted bilayer \(\mathrm{WSe_2}\)},
  author={Li, En and Hu, Jin-Xin and Feng, Xuemeng and Zhou, Zishu and An, Liheng and Law, Kam Tuen and Wang, Ning and Lin, Nian},
  journal={Nat. Commun.},
  volume={12},
  number={1},
  pages={1--7},
  year={2021},
  publisher={Nature Publishing Group}
}

@article{shabani2021deep,
  title={Deep moir{\'e} potentials in twisted transition metal dichalcogenide bilayers},
  author={Shabani, Sara and Halbertal, Dorri and Wu, Wenjing and Chen, Mingxing and Liu, Song and Hone, James and Yao, Wang and Basov, Dmitri N and Zhu, Xiaoyang and Pasupathy, Abhay N},
  journal={Nat. Phys.},
  volume={17},
  number={6},
  pages={720--725},
  year={2021},
  publisher={Nature Publishing Group}
}

@article{li2021imaging2,
  title={Imaging moir{\'e} flat bands in three-dimensional reconstructed \(\mathrm{WSe_2}\)/\(\mathrm{WS_2}\) superlattices},
  author={Li, Hongyuan and Li, Shaowei and Naik, Mit H and Xie, Jingxu and Li, Xinyu and Wang, Jiayin and Regan, Emma and Wang, Danqing and Zhao, Wenyu and Zhao, Sihan and others},
  journal={Nat. Mater.},
  volume={20},
  number={7},
  pages={945--950},
  year={2021},
  publisher={Nature Publishing Group}
}

@article{dai2016twisted,
  title={Twisted bilayer graphene: Moir{\'e} with a twist},
  author={Dai, Shuyang and Xiang, Yang and Srolovitz, David J},
  journal={Nano Lett.},
  volume={16},
  number={9},
  pages={5923--5927},
  year={2016},
  publisher={ACS Publications}
}

@article{bai2020excitons,
  title={Excitons in strain-induced one-dimensional moir{\'e} potentials at transition metal dichalcogenide heterojunctions},
  author={Bai, Yusong and Zhou, Lin and Wang, Jue and Wu, Wenjing and McGilly, Leo J and Halbertal, Dorri and Lo, Chiu Fan Bowen and Liu, Fang and Ardelean, Jenny and Rivera, Pasqual and others},
  journal={Nat. Mater.},
  volume={19},
  number={10},
  pages={1068--1073},
  year={2020},
  publisher={Nature Publishing Group}
}

@article{latychevskaia2018convergent,
  title={Convergent beam electron holography for analysis of van der Waals heterostructures},
  author={Latychevskaia, Tatiana and Woods, Colin Robert and Wang, Yi Bo and Holwill, Matthew and Prestat, Eric and Haigh, Sarah J and Novoselov, Kostya S},
  journal={Proc. Natl. Acad. Sci.},
  volume={115},
  number={29},
  pages={7473--7478},
  year={2018},
  publisher={National Acad Sciences}
}

@article{sung2022torsional,
  title={Torsional Periodic Lattice Distortions and Diffraction of Twisted 2D Materials},
  author={Sung, Suk Hyun and Goh, Yin Min and Yoo, Hyobin and Engelke, Rebecca and Xie, Hongchao and Zhang, Kuan and Li, Zidong and Ye, Andrew and Deotare, Parag B and Tadmor, Ellad B and others},
  journal={arXiv preprint arXiv:2203.06510},
  year={2022}
}

@article{pelz2017low,
  title={Low-dose cryo electron ptychography via non-convex Bayesian optimization},
  author={Pelz, Philipp Michael and Qiu, Wen Xuan and B{\"u}cker, Robert and Kassier, G{\"u}nther and Miller, RJ Dwayne},
  journal={Scientific reports},
  volume={7},
  number={1},
  pages={9883},
  year={2017},
  publisher={Nature Publishing Group UK London}
}

@article{ophus2014recording,
  title={Recording and using 4D-STEM datasets in materials science},
  author={Ophus, Colin and Ercius, Peter and Sarahan, Michael and Czarnik, Cory and Ciston, Jim},
  journal={Microscopy and Microanalysis},
  volume={20},
  number={S3},
  pages={62--63},
  year={2014},
  publisher={Cambridge University Press}
}

@article{chen2020mixed,
  title={Mixed-state electron ptychography enables sub-angstrom resolution imaging with picometer precision at low dose},
  author={Chen, Zhen and Odstrcil, Michal and Jiang, Yi and Han, Yimo and Chiu, Ming-Hui and Li, Lain-Jong and Muller, David A},
  journal={Nature communications},
  volume={11},
  number={1},
  pages={2994},
  year={2020},
  publisher={Nature Publishing Group UK London}
}

@article{jiang2018electron,
  title={Electron ptychography of 2D materials to deep sub-{\aa}ngstr{\"o}m resolution},
  author={Jiang, Yi and Chen, Zhen and Han, Yimo and Deb, Pratiti and Gao, Hui and Xie, Saien and Purohit, Prafull and Tate, Mark W and Park, Jiwoong and Gruner, Sol M and others},
  journal={Nature},
  volume={559},
  number={7714},
  pages={343--349},
  year={2018},
  publisher={Nature Publishing Group UK London}
}

@article{yang2024local,
  title={Local-orbital ptychography for ultrahigh-resolution imaging},
  author={Yang, Wenfeng and Sha, Haozhi and Cui, Jizhe and Mao, Liangze and Yu, Rong},
  journal={Nature Nanotechnology},
  pages={1--6},
  year={2024},
  publisher={Nature Publishing Group UK London}
}

@article{van2023rotational,
  title={Rotational and dilational reconstruction in transition metal dichalcogenide moir{\'e} bilayers},
  author={Van Winkle, Madeline and Craig, Isaac M and Carr, Stephen and Dandu, Medha and Bustillo, Karen C and Ciston, Jim and Ophus, Colin and Taniguchi, Takashi and Watanabe, Kenji and Raja, Archana and others},
  journal={Nature communications},
  volume={14},
  number={1},
  pages={2989},
  year={2023},
  publisher={Nature Publishing Group UK London}
}

@article{craig2024local,
  title={Local atomic stacking and symmetry in twisted graphene trilayers},
  author={Craig, Isaac M and Van Winkle, Madeline and Groschner, Catherine and Zhang, Kaidi and Dowlatshahi, Nikita and Zhu, Ziyan and Taniguchi, Takashi and Watanabe, Kenji and Griffin, Sin{\'e}ad M and Bediako, D Kwabena},
  journal={Nature Materials},
  pages={1--8},
  year={2024},
  publisher={Nature Publishing Group UK London}
}

@article{scherzer1949theoretical,
  title={The theoretical resolution limit of the electron microscope},
  author={Scherzer, O},
  journal={Journal of Applied Physics},
  volume={20},
  number={1},
  pages={20--29},
  year={1949},
  publisher={American Institute of Physics}
}

@article{hytch1998quantitative,
  title={Quantitative measurement of displacement and strain fields from HREM micrographs},
  author={H{\"y}tch, MJ and Snoeck, E and Kilaas, R},
  journal={Ultramicroscopy},
  volume={74},
  number={3},
  pages={131--146},
  year={1998},
  publisher={Elsevier}
}

@article{humphry2012ptychographic,
  title={Ptychographic electron microscopy using high-angle dark-field scattering for sub-nanometre resolution imaging},
  author={Humphry, MJ and Kraus, B and Hurst, AC and Maiden, AM and Rodenburg, JM},
  journal={Nature communications},
  volume={3},
  number={1},
  pages={730},
  year={2012},
  publisher={Nature Publishing Group UK London}
}

@article{gabor,
  title={A New Microscopic Principle.},
  author={Gabor, D},
  journal={Nature},
  volume={161},
  pages={777--778},
  year={1948},
}

@article{hegerl1970dynamische,
  title={Dynamische theorie der kristallstrukturanalyse durch elektronenbeugung im inhomogenen prim{\"a}rstrahlwellenfeld},
  author={Hegerl, Reiner and Hoppe, Walter},
  journal={Berichte der Bunsengesellschaft f{\"u}r physikalische Chemie},
  volume={74},
  number={11},
  pages={1148--1154},
  year={1970},
  publisher={Wiley Online Library}
}

@article{de2022imaging,
  title={Imaging moir{\'e} deformation and dynamics in twisted bilayer graphene},
  author={de Jong, Tobias A and Benschop, Tjerk and Chen, Xingchen and Krasovskii, Eugene E and de Dood, Michiel JA and Tromp, Rudolf M and Allan, Milan P and Van der Molen, Sense Jan},
  journal={Nature Communications},
  volume={13},
  number={1},
  pages={70},
  year={2022},
  publisher={Nature Publishing Group UK London}
}

@article{tate2016high,
  title={High dynamic range pixel array detector for scanning transmission electron microscopy},
  author={Tate, Mark W and Purohit, Prafull and Chamberlain, Darol and Nguyen, Kayla X and Hovden, Robert and Chang, Celesta S and Deb, Pratiti and Turgut, Emrah and Heron, John T and Schlom, Darrell G and others},
  journal={Microscopy and Microanalysis},
  volume={22},
  number={1},
  pages={237--249},
  year={2016},
  publisher={Oxford University Press}
}

@article{suzuki2016dark,
  title={Dark-field X-ray ptychography: Towards high-resolution imaging of thick and unstained biological specimens},
  author={Suzuki, Akihiro and Shimomura, Kei and Hirose, Makoto and Burdet, Nicolas and Takahashi, Yukio},
  journal={Scientific reports},
  volume={6},
  number={1},
  pages={35060},
  year={2016},
  publisher={Nature Publishing Group UK London}
}

@article{rodenburg1993experimental,
  title={Experimental tests on double-resolution coherent imaging via STEM},
  author={Rodenburg, JM and McCallum, BC and Nellist, PD},
  journal={Ultramicroscopy},
  volume={48},
  number={3},
  pages={304--314},
  year={1993},
  publisher={Elsevier}
}

@Inbook{Rodenburg2019,
author="Rodenburg, John
and Maiden, Andrew",
editor="Hawkes, Peter W.
and Spence, John C. H.",
title="Ptychography",
bookTitle="Springer Handbook of Microscopy",
year="2019",
publisher="Springer International Publishing",
address="Cham",
pages="819--904",
isbn="978-3-030-00069-1",
doi="10.1007/978-3-030-00069-1_17",
url="https://doi.org/10.1007/978-3-030-00069-1_17"
}

@article{ruska1931magnetische,
  title={Die magnetische Sammelspule f{\"u}r schnelle Elektronenstrahlen},
  author={Ruska, Ernst and Knoll, Max},
  journal={The magnetic concentrating coil for fast electron beams.) Z. techn. Physik},
  volume={12},
  pages={389--400},
  year={1931}
}

@article{hruszkewycz2017high,
  title={High-resolution three-dimensional structural microscopy by single-angle Bragg ptychography},
  author={Hruszkewycz, Stephan O and Allain, Marc and Holt, Martin V and Murray, Conal E and Holt, Judson R and Fuoss, Paul H and Chamard, Virginie},
  journal={Nature materials},
  volume={16},
  number={2},
  pages={244--251},
  year={2017},
  publisher={Nature Publishing Group UK London}
}

@article{pfeiffer2018x,
  title={X-ray ptychography},
  author={Pfeiffer, Franz},
  journal={Nature Photonics},
  volume={12},
  number={1},
  pages={9--17},
  year={2018},
  publisher={Nature Publishing Group UK London}
}

@article{takahashi2013bragg,
  title={Bragg x-ray ptychography of a silicon crystal: Visualization of the dislocation strain field and the production of a vortex beam},
  author={Takahashi, Yukio and Suzuki, Akihiro and Furutaku, Shin and Yamauchi, Kazuto and Kohmura, Yoshiki and Ishikawa, Tetsuya},
  journal={Physical Review B—Condensed Matter and Materials Physics},
  volume={87},
  number={12},
  pages={121201},
  year={2013},
  publisher={APS}
}

@article{kim2018three,
  title={Three-dimensional imaging of phase ordering in an Fe-Al alloy by Bragg ptychography},
  author={Kim, Chan and Chamard, Virginie and Hallmann, J{\"o}rg and Roth, Thomas and Lu, Wei and Boesenberg, Ulrike and Zozulya, Alexey and Leake, Steven and Madsen, Anders},
  journal={Physical Review Letters},
  volume={121},
  number={25},
  pages={256101},
  year={2018},
  publisher={APS}
}

@article{zhang_anomalous_2023,
	title = {Anomalous {Interfacial} {Electron}-{Transfer} {Kinetics} in {Twisted} {Trilayer} {Graphene} {Caused} by {Layer}-{Specific} {Localization}},
	volume = {9},
	journal = {ACS Central Science},
	author = {Zhang, Kaidi and Yu, Yun and Carr, Stephen and Babar, Mohammad and Zhu, Ziyan and Kim, Bryan Junsuh and Groschner, Catherine and Khaloo, Nikta and Taniguchi, Takashi and Watanabe, Kenji and Viswanathan, Venkatasubramanian and Bediako, D. Kwabena},
	year = {2023},
	pages = {1119--1128},
}

@article{yu_tunable_2022,
	title = {Tunable angle-dependent electrochemistry at twisted bilayer graphene with moiré flat bands},
	volume = {14},
	number = {3},
	journal = {Nature Chemistry},
	author = {Yu, Yun and Zhang, Kaidi and Parks, Holden and Babar, Mohammad and Carr, Stephen and Craig, Isaac M. and Van Winkle, Madeline and Lyssenko, Artur and Taniguchi, Takashi and Watanabe, Kenji and Viswanathan, Venkatasubramanian and Bediako, D. Kwabena},
	year = {2022},
	pages = {267--273},
}

@article{yang2022tunable,
	Author = {Xu, Yang and Kang, Kaifei and Watanabe, Kenji and Taniguchi, Takashi and Mak, Kin Fai and Shan, Jie},
	Journal = {Nature Nanotechnology},
	Number = {9},
	Pages = {934--939},
	Title = {A tunable bilayer Hubbard model in twisted WSe2},
	Volume = {17},
	Year = {2022}}

@article{CoelloEscalante2024TBG,
author = {Coello Escalante, Leonardo and Limmer, David T.},
title = {Microscopic Origin of Twist-Dependent Electron Transfer Rate in Bilayer Graphene},
journal = {Nano Letters},
volume = {24},
number = {46},
pages = {14868-–14874},
year = {2024},
}

@article{Maroo2026electronic,
	Author = {Maroo, Sonal and Coello Escalante, Leonardo and Wang, Yizhe and Erodici, Matthew P. and Nessralla, Jonathon N. and Tabo, Ayana and Taniguchi, Takashi and Watanabe, Kenji and Xu, Ke and Limmer, David T. and Bediako, D. Kwabena},
	Journal = {Nature},
	Number = {8113},
	Pages = {98--103},
	Title = {Electronic origin of reorganization energy in interfacial electron transfer},
	Volume = {653},
	Year = {2026}}

@article{vanwinkle2025nanoscale,
	Author = {Van Winkle, Madeline and Zhang, Kaidi and Bediako, D. Kwabena},
	Journal = {Accounts of Chemical Research},
	Month = {02},
	Number = {3},
	Pages = {415--427},
	Title = {Nanoscale Structure and Interfacial Electrochemical Reactivity of Moir{\'e}-Engineered Atomic Layers},
	Volume = {58},
	Year = {2025}}

@article{wijk2015relaxation,
	Author = {Wijk, M M van and Schuring, A and Katsnelson, M I and Fasolino, A},
	Number = {3},
	Pages = {034010},
	Title = {Relaxation of moir{\'e}patterns for slightly misaligned identical lattices: graphene on graphite},
	Volume = {2},
	Year = {2015}}

@article{nam2017lattice,
	Author = {Nam, Nguyen N. T. and Koshino, Mikito},
	Journal = {Physical Review B},
	Month = {08},
	Number = {7},
	Pages = {075311--},
	Title = {Lattice relaxation and energy band modulation in twisted bilayer graphene},
	Volume = {96},
	Year = {2017}}

@article{tadmor2018structural,
	Author = {Zhang, Kuan and Tadmor, Ellad B.},
	Journal = {Journal of the Mechanics and Physics of Solids},
	Pages = {225--238},
	Title = {Structural and electron diffraction scaling of twisted graphene bilayers},
	Volume = {112},
	Year = {2018}}

@article{alden2013strain,
	Author = {Alden, Jonathan S. and Tsen, Adam W. and Huang, Pinshane Y. and Hovden, Robert and Brown, Lola and Park, Jiwoong and Muller, David A. and McEuen, Paul L.},
	Journal = {Proceedings of the National Academy of Sciences},
	Pages = {11256--11260},
	Title = {Strain solitons and topological defects in bilayer graphene},
	Volume = {110},
	Year = {2013}}

@article{woods2014commensurate,
	Author = {Woods, C. R. and Britnell, L. and Eckmann, A. and Ma, R. S. and Lu, J. C. and Guo, H. M. and Lin, X. and Yu, G. L. and Cao, Y. and Gorbachev, R. V. and Kretinin, A. V. and Park, J. and Ponomarenko, L. A. and Katsnelson, M. I. and Gornostyrev, Yu. N. and Watanabe, K. and Taniguchi, T. and Casiraghi, C. and Gao, H-J. and Geim, A. K. and Novoselov, K. S.},
	Journal = {Nature Physics},
	Number = {6},
	Pages = {451--456},
	Title = {Commensurate--incommensurate transition in graphene on hexagonal boron nitride},
	Volume = {10},
	Year = {2014}}

@article{uri2020mapping,
	Author = {Uri, A. and Grover, S. and Cao, Y. and Crosse, J. A. and Bagani, K. and Rodan-Legrain, D. and Myasoedov, Y. and Watanabe, K. and Taniguchi, T. and Moon, P. and Koshino, M. and Jarillo-Herrero, P. and Zeldov, E.},
	Journal = {Nature},
	Number = {7806},
	Pages = {47--52},
	Title = {Mapping the twist-angle disorder and Landau levels in magic-angle graphene},
	Volume = {581},
	Year = {2020}}

@article{wilson2020disorder,
	Author = {Wilson, Justin H. and Fu, Yixing and Das Sarma, S. and Pixley, J. H.},
	Journal = {Physical Review Research},
	Month = {06},
	Number = {2},
	Pages = {023325--},
	Title = {Disorder in twisted bilayer graphene},
	Volume = {2},
	Year = {2020}}

@article{huder2018electronic,
	Author = {Huder, Lo{\"\i}c and Artaud, Alexandre and Le Quang, Toai and de Laissardi{\`e}re, Guy Trambly and Jansen, Aloysius G. M. and Lapertot, G{\'e}rard and Chapelier, Claude and Renard, Vincent T.},
	Journal = {Physical Review Letters},
	Month = {04},
	Number = {15},
	Pages = {156405--},
	Title = {Electronic Spectrum of Twisted Graphene Layers under Heterostrain},
	Volume = {120},
	Year = {2018}}

@article{bi2019designing,
	Author = {Bi, Zhen and Yuan, Noah F. Q. and Fu, Liang},
	Journal = {Physical Review B},
	Month = {07},
	Number = {3},
	Pages = {035448--},
	Title = {Designing flat bands by strain},
	Volume = {100},
	Year = {2019}}

@article{craig2024considerations,
    author = {Craig, Isaac M. and Van Winkle, Madeline and Ophus, Colin and Bediako, D. Kwabena},
    title = {Considerations for extracting moiré-level strain from dark field intensities in transmission electron microscopy},
    journal = {Journal of Applied Physics},
    volume = {136},
    number = {7},
    pages = {074301},
    year = {2024},
}

@article{scherzer1936,
	Author = {Scherzer, O. },
	Journal = {Zeitschrift f{\"u}r Physik},
	Number = {9},
	Pages = {593--603},
	Title = {{\"U}ber einige Fehler von Elektronenlinsen},
	Volume = {101},
	Year = {1936}}

@article{chen2021electron,
	Author = {Chen, Zhen and Jiang, Yi and Shao, Yu-Tsun and Holtz, Megan E. and Odstr{\v c}il, Michal and Guizar-Sicairos, Manuel and Hanke, Isabelle and Ganschow, Steffen and Schlom, Darrell G. and Muller, David A.},
	Journal = {Science},
	Month = {2026/07/04},
	Number = {6544},
	Pages = {826--831},
	Title = {Electron ptychography achieves atomic-resolution limits set by lattice vibrations},
	Volume = {372},
	Year = {2021}}

@article{pennycook2015efficient,
	Author = {Pennycook, Timothy J. and Lupini, Andrew R. and Yang, Hao and Murfitt, Matthew F. and Jones, Lewys and Nellist, Peter D.},
	Journal = {Ultramicroscopy},
	Pages = {160--167},
	Title = {Efficient phase contrast imaging in STEM using a pixelated detector. Part 1: Experimental demonstration at atomic resolution},
	Volume = {151},
	Year = {2015}}

@article{gao2017electron,
	Author = {Gao, Si and Wang, Peng and Zhang, Fucai and Martinez, Gerardo T. and Nellist, Peter D. and Pan, Xiaoqing and Kirkland, Angus I.},
	Journal = {Nature Communications},
	Number = {1},
	Pages = {163},
	Title = {Electron ptychographic microscopy for three-dimensional imaging},
	Volume = {8},
	Year = {2017}}

@article{Pasztor2019Holographic,
	Author = {P{\'a}sztor, {\'A}rp{\'a}d and Scarfato, Alessandro and Spera, Marcello and Barreteau, C{\'e}line and Giannini, Enrico and Renner, Christoph},
	Journal = {Physical Review Research},
	Journal1 = {Phys. Rev. Res.},
	Month = {11},
	Number = {3},
	Pages = {033114--},
	Title = {Holographic imaging of the complex charge density wave order parameter},
	Volume = {1},
	Year = {2019},
	}

@article{Pasztor2021multiband,
	Author = {P{\'a}sztor, {\'A}rp{\'a}d and Scarfato, Alessandro and Spera, Marcello and Flicker, Felix and Barreteau, C{\'e}line and Giannini, Enrico and Wezel, Jasper van and Renner, Christoph},
	Journal = {Nature Communications},
	Number = {1},
	Pages = {6037},
	Title = {Multiband charge density wave exposed in a transition metal dichalcogenide},
	Volume = {12},
	Year = {2021}}

@article{pasztor2024delusive,
	Author = {P{\'a}sztor, {\'A}rp{\'a}d and Pushkarna, Ishita and Renner, Christoph},
	Number = {3},
	Pages = {035017},
	Title = {Delusive chirality and periodic strain pattern in moir{\'e}systems},
	Volume = {11},
	Year = {2024}
    }

@article{ke2022moire,
	Author = {Ke, Xiaoxing and Zhang, Manchen and Zhao, Kangning and Su, Dong},
	Journal = {Small Methods},
	Month = {2026/07/04},
	Number = {1},
	Pages = {2101040},
	Title = {Moir{\'e}Fringe Method via Scanning Transmission Electron Microscopy},
	Volume = {6},
	Year = {2022}}

@article{faruqi2011electronic,
	Author = {Faruqi, A. R. and McMullan, G.},
	Number = {3},
	Pages = {357-390},
	Title = {Electronic detectors for electron microscopy},
	Volume = {44},
	Year = {2011}}

@article{levin2021direct,
	Author = {Levin, Barnaby D A},
	Number = {4},
	Pages = {042005},
	Title = {Direct detectors and their applications in electron microscopy for materials science},
	Volume = {4},
	Year = {2021}}

@article{Hoppe1969,
	Author = {Hoppe, W.},
	Journal = {Acta Crystallographica Section A},
	Month = {Jul},
	Number = {4},
	Pages = {495--501},
	Title = {{Beugung im inhomogenen Prim{\"{a}}rstrahlwellenfeld. I. Prinzip einer Phasenmessung von Elektronenbeungungsinterferenzen}},
	Volume = {25},
	Year = {1969}}

@misc{clark2025electronptychography,
      title={Electron Ptychography}, 
      author={L. Clark and P. D. Nellist},
      year={2025},
      eprint={2503.10917},
      archivePrefix={arXiv},
      primaryClass={physics.optics},
      url={https://arxiv.org/abs/2503.10917}, 
}

@article{rodenburg2025ptychography,
	Author = {Rodenburg, John},
	Journal = {Journal of Microscopy},
	Month = {2026/07/04},
	Number = {2},
	Pages = {153--155},
	Title = {Ptychography: A brief introduction},
	Volume = {300},
	Year = {2025}}

@article{miao2025computational,
	Author = {Miao, Jianwei},
	Journal = {Nature},
	Number = {8045},
	Pages = {281--295},
	Title = {Computational microscopy with coherent diffractive imaging and ptychography},
	Volume = {637},
	Year = {2025}}

@article{zheng2021concept,
	Author = {Zheng, Guoan and Shen, Cheng and Jiang, Shaowei and Song, Pengming and Yang, Changhuei},
	Journal = {Nature Reviews Physics},
	Number = {3},
	Pages = {207--223},
	Title = {Concept, implementations and applications of Fourier ptychography},
	Volume = {3},
	Year = {2021}}

@article{wang2023optical,
	Author = {Wang, Tianbo and Jiang, Shaowei and Song, Pengming and Wang, Ruihai and Yang, Liming and Zhang, Terrance and Zheng, Guoan},
	Journal = {Biomedical Optics Express},
	Number = {2},
	Pages = {489--532},
	Title = {Optical ptychography for biomedical imaging: recent progress and future directions {$[$}Invited{$]$}},
	Volume = {14},
	Year = {2023}}

@article{jain2017structure,
	Author = {Jain, Sandeep K and Juri{\v c}i{\'c}, Vladimir and Barkema, Gerard T},
	Number = {1},
	Pages = {015018},
	Title = {Structure of twisted and buckled bilayer graphene},
	Volume = {4},
	Year = {2017}}

@article{park2022robust,
	Author = {Park, Jeong Min and Cao, Yuan and Xia, Li-Qiao and Sun, Shuwen and Watanabe, Kenji and Taniguchi, Takashi and Jarillo-Herrero, Pablo},
	Journal = {Nature Materials},
	Number = {8},
	Pages = {877--883},
	Title = {Robust superconductivity in magic-angle multilayer graphene family},
	Volume = {21},
	Year = {2022}}

@misc{fujimoto2025four,
      title={Four Moir\'e materials at One Magic Angle in Helical Quadrilayer Graphene}, 
      author={Manato Fujimoto and Naoto Nakatsuji and Ashvin Vishwanath and Patrick Ledwith},
      year={2025},
      eprint={2510.02444},
      archivePrefix={arXiv},
      primaryClass={cond-mat.str-el},
}

@article{Han2024correlated,
	Author = {Han, Tonghang and Lu, Zhengguang and Scuri, Giovanni and Sung, Jiho and Wang, Jue and Han, Tianyi and Watanabe, Kenji and Taniguchi, Takashi and Park, Hongkun and Ju, Long},
	Journal = {Nature Nanotechnology},
	Number = {2},
	Pages = {181--187},
	Title = {Correlated insulator and Chern insulators in pentalayer rhombohedral-stacked graphene},
	Volume = {19},
	Year = {2024}}

@article{yu2022tuning,
	Author = {Yu, Yun and Van Winkle, Madeline and Bediako, D. Kwabena},
	Journal = {Trends in Chemistry},
	Number = {10},
	Pages = {857--859},
	Title = {Tuning interfacial chemistry with twistronics},
	Volume = {4},
	Year = {2022}}

\end{document}